\documentclass[useAMS,usenatbib]{mn2e}
\usepackage{times}

\usepackage{graphics}
\usepackage{graphicx}

\title[Hot, Dense Matter Accretion onto a Black Hole]{Hot, Dense Matter Accretion onto a Black Hole}
\author[M. Yokosawa, S. Uematsu and J. Abe]{M.~Yokosawa,$^{1}$\thanks{E-mail:
yokosawa@mx.ibaraki.ac.jp (MY)} S.~Uematsu$^1$ and J.~Abe$^1$\\
$^{1}$Department of Astrophysics, Faculty of Natural Science, Ibaraki University, Mito Bunkyo 2-1-1 Post Num. 310-8512, Japan}
\begin{document}


\pagerange{\pageref{firstpage}--\pageref{lastpage}} \pubyear{2004}

\maketitle

\label{firstpage}

\begin{abstract}
The accretion of hot, dense matter which consists of heavy nuclei, free
 nucleons, degenerated electrons, photons and neutrinos is studied. The
 composition of free nucleons and their chemical potentials are provided
 through the equation of state for hot, dense matter proposed by
 \citet{LS}. The numbers of 
 leptons are calculated through the reactions of neutrinos and through
 the simplified transfers of neutrinos in the 
 accreting matter. The thermally equilibrium fluid structure of the
 accretion is solved. The angular 
 momentum transfer in the disk is analyzed with the shear stress
 which is driven in the
 frame of the general relativity. When the mass of 
 a central black hole and the accretion rate are selected as  $M_{BH}
 \approx 3{\rm M}_\odot$ and $\dot{M} \approx 0.1{\rm 
 M}_\odot$sec$^{-1}$, which provide 
 the typical luminosity of gamma ray bursts(GRBs), the fraction of free
 protons in the accreting matter becomes very small, $Y_p \approx
 10^{-4}$, while that of free neutrons is closer to unity, $Y_n
 \approx 0.7$. Then the  antielectron neutrinos
 $\bar{\nu}_e$  can freely escape through the disk but the
 electron neutrinos $\nu_e$ are almost absorbed in the disk.  Thus the
 frequent collisions of $\bar{\nu}_e$ with $\nu_e$ over the
 disk couldn't be occurred. The
 accretion disk is cooled mainly by $\bar{\nu}_e$, which suppresses
 the increase of temperature and increases the density in 
 the accreting matter such as $T \approx 3\times 10^{10}$K
 and $\rho \approx 3\times 10^{13}$g/cm$^3$ at the inner side of the disk. The scattering
 optical depths of $\bar{\nu}_e$ and $\nu_e$ then reach to be very large, 
 $\tau_s(\bar{\nu}_e) \approx \tau_s({\nu}_e) \approx 10^2$. Thus the 
 accretion disk could be thermally unstable within the duration of
 diffusion time of neutrinos, $t_{diff} \approx 10$(ms). The ram pressure produced by many scattering of $\bar{\nu}_e$ are
 very strong, which might produce the neutrino driven wind or
 jets. The luminosity and mean energy of neutrinos, $L_\nu$,
 $\bar{E}_\nu$, ejected from the disk increases with the specific
 angular momentum of a black hole. The mean energy is
 inversely proportional to the central mass, $\bar{E}_\nu \propto M_{BH}^{-1}$, while the luminosity $L_\nu$ is
 independent of its mass, $ M_{BH}$. The diagram determining the physical
 properties of a central  black hole from the observation of each flavor of neutrino is proposed.
\end{abstract}


\section{Introduction}

An intense gamma-ray burst, GRB030329, allowed us to carry out the very detailed 
observations of optical, x-ray, and radio counterparts to the burst. Its photometry 
and spectroscopy led to the separation between the afterglow and SN contributions 
\citep{MTS, KWBT03, LPKN}. The SN component is similar to SN 1998bw; an unusual Type 
Ic SN \citep{STTHKS, FRD}. It makes clear that at least some GRBs arise from 
core-collapse SNe. GRB030329 was seen as a double-pulsed burst with the
total duration time of $\sim$30 sec \citep{VND}. Its total isotropic
energy is $E_\gamma \sim 10^{52}$ erg. The GRBs with known redshifts
distribute the emission energy $E_\gamma$ over the
wide range of energy such as $10^{48}$ erg $\ge E_\gamma \ge 
10^{54}$erg\citep{JG}. The duration times of GRBs extend 
from $\sim$10ms to $\sim 10^2$ sec\citep{P99,ST01,LR02}. The light curves of GRBs show really various profiles. However, to fully 
understand the emission mechanisms of the observed $\gamma$-ray bursts and of the 
afterglows one must have a firm handle on the generation process of the $\gamma$-ray 
bust.

The fireball model of the $\gamma$-ray bust has succeeded in understanding the 
characteristics of some light profiles of GRBs \citep{P99}. Its model was proposed 
to solve the compactness problem, that is, to understand both the short time variation 
of $\gamma$-ray radiation and the optically thin process in the emitting region of 
the $\gamma$-ray photons \citep{GFR, CK}. The solution of the fireball model requires 
the beaming jets with the emitting region relativistically moving to our direction. 
The two processes are proposed for the formation of the jets, one is the 
magnetohydrodynamical process \citep{TCQ, LB}, and the other is the neutrino process 
\citep{PWF,NPK,KM,MPN}. Further more the two different roles are expected for the 
neutrinos to form the jets. The frequent collisions between neutrinos and 
antineutrinos produce the fireball with light elements which expands 
relativistically in a funnel around a rotation axis. The other role is the 
acceleration of plasmas by the collisions of neutrinos with nucleons or electrons, 
which produces the neutrino driven jets. The sufficient collisions between $\nu$ and 
$\bar\nu$ at the central region of the accretion disk were proposed by \citet{PWF}. 
They assumed the accretion disk is optically thin for the transfer of neutrinos. But, 
\citet{MPN} showed that the accretion disk becomes optically thick for neutrino 
transfer and then the neutrinos trapped in the matter fall into a black hole. 

However, the composition of accreting matter should be carefully
considered. The production rates of neutrinos and their opacities directly
depend on the density of free nucleons. \citet{KM} were thought in
question whether the neutrinos can cool the accretion disk. They treated
the accreting material in $\beta$ equilibrium and showed that there
exist just a small number of free nucleons and then the neutrino processes
are no longer effective in cooling of the disk. This
result is in a striking contrast to the previous works on the accretion
disk in which the material almost consists of free
nucleons produced by the photodisintegration process and the dominant
cooling was carried out by neutrinos. Material at high
temperature $T \ge 5\times10^9$K becomes in nuclear statistical
equilibrium. Its composition of material depends on the number ratio of
photon to baryon, $\psi = n_\gamma /\mu_H n_B$, where $\mu_H$ is the mean
molecular weight, $n_\gamma$ and $n_B$ are the number densities of photons and
baryons \citep{MMHWH}. If $\psi$ is large, $\psi \gg 1$, the baryons form
a dissociated gas of nucleons and alpha particles. However, if $\psi$ is
of order unity or less, iron-group nuclei are typically favored. The
ratio $\psi$ is expressed as $\psi=0.034T_{11}^3/\rho_{12}$ in terms of
the temperature $T_{11}$ in units of $10^{11}$K and the density
$\rho_{12}$ in units of $10^{12}$g/cm$^3$. In the case of a typical
accretion with $\dot M=0.1{\rm M}_\odot$/sec and $M_{BH}=3{\rm M}_\odot$, the temperature and the density become
are $10^{10} < T < 10^{11}$(K) and $10^{11} < \rho < 10^{13}$(g/cm$^3$) at
the inner region of the accretion disk. Then the values of $\psi$ become as
$\psi=3.4\times 10^{-2 \sim -3}$. One should treat the accreting matter
with a large abundance of heavy nuclei. We determine the densities of
free neutron and proton, alpha particle, and heavy nuclei by using the
equation of state for hot, dense matter given by \citet{LS}.
The changes of lepton number per a
baryon, $\dot{Y}_l$, are calculated through the reactions of neutrinos
with these baryons.

The thermal equilibrium structure of the accretion disk should be determined in the frame of a general relativity. \citet{MW} showed
that the core-collapse in the evolved star with the main sequence mass
, $M \ge 35{\rm M}_\odot$, forms a black hole at the center with
the rotating disk balanced with gravity when the evolved star has the some
large specific angular momentum. Then the specific angular momentum parameter 
$a$ of the falling matter into a black hole becomes larger to be $a \ge
0.9$. The falling
velocity of the rotating disk is required to be very small in comparison with
light velocity, since the time duration of some GRBs is a few seconds, i.e., $v_{fall} \approx scale of disk/duration time \approx 10^2r_g/a few sec \approx 10^{-3}c$,
where $r_g$ is the gravitational radius of a black hole and $c$ is the light velocity. The accretion disk with slow falling velocity should be investigated in the frame of rotating spacetime. The heating rate due to viscosity is proportional to the shear
stress $\sigma_{r\varphi}$. In the static spacetime its stress
increases with the differential rotation of the accreting flow
$\sigma_{r\varphi} \propto \partial_r \Omega$, where $\Omega$ is the
angular velocity of the falling flow. In the rotating spacetime the
differential rotation of the fluid observed in the frame
co-rotating with spacetime is physically meaningful \citep{TPM}. In the extreme
Kerr the relative angular velocity $\Omega -\omega$ has the maximum at
the radius $r \approx 2r_g$ when $\Omega$ is the angular velocity of the Keplerian
orbit and $\omega$ is the angular velocity of
spacetime. It is convenient for the analysis of energy balance
and of the transfer of angular momentum to introduce the orbiting frame
with orthogonal tetrad which rotates with Keplarian velocity. In this paper the shear stress $\sigma_{r\varphi}$ and the transfer of angular momentum are solved by using the orbiting frame.
 
The power by advection contributes seriously to the energy balance in  
the accretion disk. The previous works on the accretion disk with
neutrino loss had  
been studied in the frame of an advection-dominated accretion flow
\citep{MPN,KM,NPK,PWF}. The role of the advection in the energy
balance depends on the velocity profile since the power 
by the advection is expressed as $P\vec{\nabla}\cdot\vec{v}$, where $P$ and 
$\vec{v}$ are the pressure and velocity of the flow. When
$\vec{\nabla}\cdot\vec{v}  
> 0$, the accreting matter is cooled by the expansion of fluid. On the
other hand when $\vec{\nabla}\cdot\vec{v} < 0$, it is heated by the
compression like as in the core-collapse 
in a supernova explosion. The radial velocity profile of the accreting flow is
determined by the energy balance of the disk and by the transfer of angular
momentum. The scale height at the inner portion of the disk with the
temperature $T=10^{10\sim 11}$K is very thin. The
angular velocity of the accretion disk is then approximated to 
be Keplerian. We determine the radial profile of the velocity by solving
the transfer of angular momentum with the thermal energy balance of the disk and evaluate the efficiency of advection in the energy balance.

The thermal stability of the accretion disk might depend on the scattering
optical depth of $\bar{\nu}_e$. The thermal energy in the disk is
carried out mainly by $\bar{\nu}_e$. The collision frequency of
$\bar{\nu}_e$ with the nucleons or electrons reaches to $N\approx \tau_{\bar\nu_e s }^2\approx 10^4$ within the
disk for the typical case of the accretion. The escaping time
from the disk is then $t_{esc}\approx \tau_{\bar\nu_e s} H/c
\approx 10$(ms), where $H$ is the height of the disk. The thermal energy produced by viscous heating stays within the disk in the time $\Delta t\approx t_{esc}$ and might induce the thermal instability. 
We shall briefly analyze the thermal instability. 
  
If the luminosity of each flavor
of neutrino and its mean energy are
observed the physics of GRBs will be cleared. We
shall calculate the luminosity $L_{\nu_i}$ and the mean energy
$\bar{E}_{\nu_i}$ for each type of neutrino $\nu_i$ 
emitted from the accretion disk. The diagram determining the mass of a central black hole, $M_{BH}$, and its specific angular momentum, $a$, from the observations of neutrinos is presented.

In this paper we study the accretion of hot dense matter onto a black
hole. In Sect. 2 we describe the equation of state for the accreting
hot, dense matter. The reactions of neutrinos 
and the changes of neutrino fractions are presented in section
2.2. The relativistic model of accretion disk with neutrino loss is given in
Sect. 3. In Sect. 4 the results of thermally equilibrium disk are
given. The profiles of compositions along the accretion flow are shown in
section 4.1. The flow structure and chemical potentials of nucleons are depicted in section 4.2. The emissivities and 
opacities of neutrinos are presented in section 4.3. The luminosity and mean energy
of neutrinos versus mass and angular momentum of a black hole are 
discussed in section 4.4.
 In Sect. 5 the thermal stability of the accretion disk is investigate. The neutrino transfer in homogeneous disk, changing rates of positrons and the dynamical properties of accretion disk are presented in Appendix. The cross sections of neutrinos, opacities, emission rates and the reaction rates of neutrinos are summarized in Tables, which are shown in Appendix.

\section{Accreting Dense Matter and Reaction of Neutrinos}
\subsection{Equation of State for Accreting Dense Matter}
It is important for the study of thermal and dynamical properties of massive accretion disks precisely to determine the fraction of free nucleons in the accreting matter. Its fraction provides the emissivities and absorptions of neutrinos. The mass fraction of free nucleons depends on the photon-to-baryon ratio $\phi$\citep{B57, MMHWH},
\begin{equation}
\phi=\frac{n_\gamma}{\rho N_A},
\end{equation}
where $N_A$ is 
Avogadro's number. We compute $\phi$ in terms 
of the temperature $T$ and density $\rho$ as 
\begin{equation}
\phi=\frac{1}{{\pi}^2} \frac{g_\gamma}{(hc/2\pi)^3} \frac{\zeta(3)(kT)^3}{\rho 
N_A}=0.034\frac{T^3_{11}}{\rho_{12}},
\end{equation}
where $g_\gamma$ is the helicity for photons. We evaluate the ratio $\phi$ for 
the neutrino dominated accreting flow (NDAF) given by \citet{MPN}. The loci of NDAF with ${\dot M} = 0.1{\rm
M}_\odot$sec$^{-1}$ and the contours of $\phi(\rho,T)$  in
the $\rho$ --- $T$ plane are shown in Fig 1. The ratio $\phi$ on the flow
is $\phi\le 0.1$, showing little number density of photons. On the other hand 
the previous works on NDAF without \citet{KM} adopted the approximate
formula for the mass fraction of free nucleons, $X_{nuc}$, \citep{QW} :
\begin{equation}
X_{nuc} = \frac{{T_{11}}^{9/8}} {\rho^{3/4}} \exp(-0.61/T_{11}).
\end{equation}
The contour of $ X_{nuc} $ is also plotted in Fig. 1. At the inner
side of NDAF its fraction, $X_{nuc}$, reaches to unity. We also  
calculated the flow profile of the standard accretion disk with the Keplerian angular 
momentum in which the fraction of free nucleons is provided through
this formula, $X_{nuc}$. The accreting velocity of the standard  
disk is much less than that of NDAF, which produces the denser profile
of the accretion flow for the same accretion rate, $\dot M = 0.1{\rm M}_\odot$sec$^{-1}$. 
Then its ratio $\phi$ is very minute, $\phi \ll 0.01$. Thus it is
necessary to handle the accreting matter with an appreciable  
abundance of heavy nuclei.

\begin{figure}
\includegraphics[width=8cm,height=6cm]{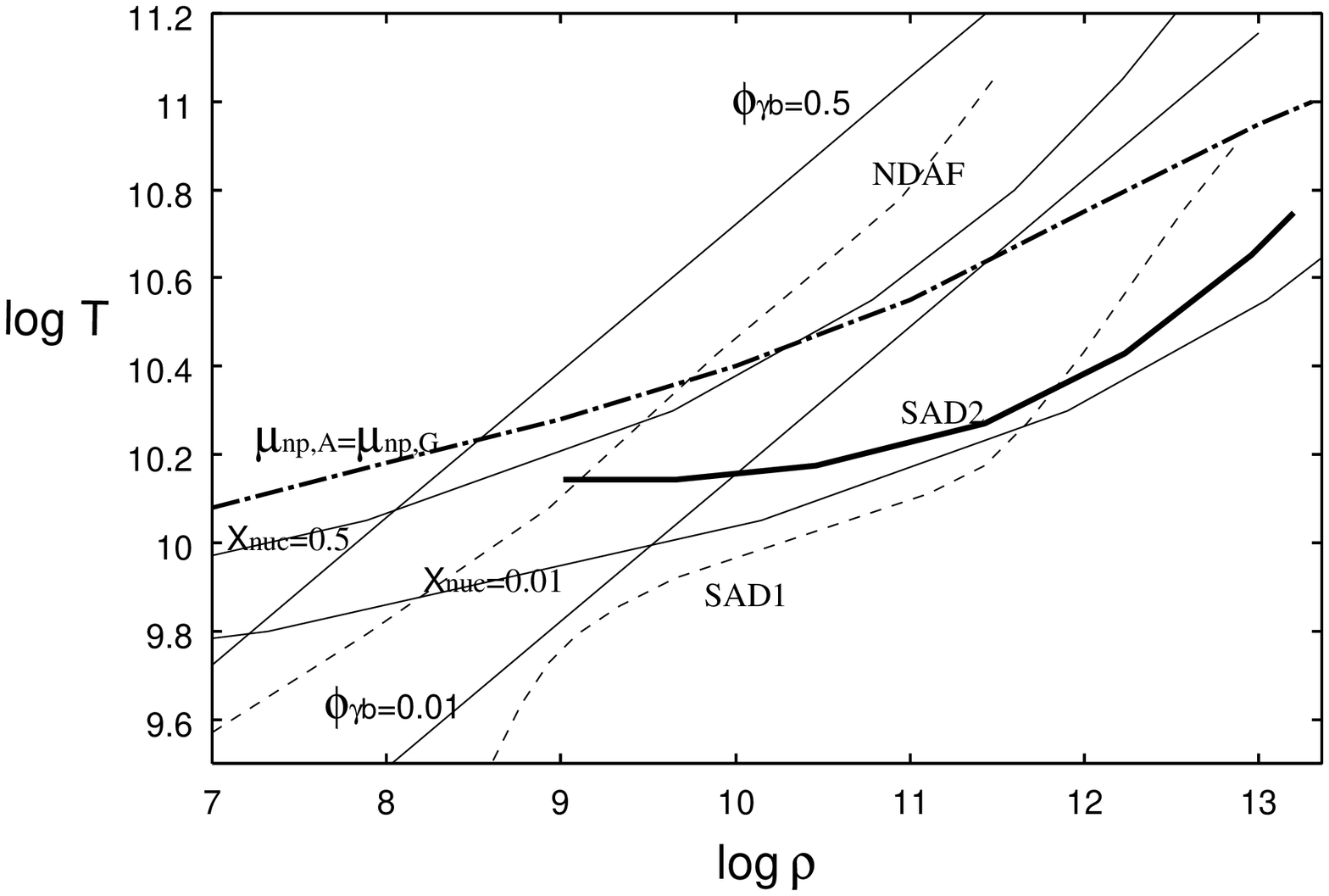}
\caption{The loci of accreting gas on the $\rho$(g cm$^{-3}$) --- $T$(K) plane. NDAF is the
 locus of the
 neutrino-dominant accretion flow with $\dot{M}=0.1{\rm M}_\odot$ sec$^{-1}$
 and $M_{BH}=3{\rm M}_\odot$. SAD1 is  the  locus of the standard accretion
 disk with Keplarian angular momentum where heavy nuclei are
ignored. SAD2 is  the  locus of the standard accretion 
 disk with heavy nuclei. The lines with the constant photon-to-baryon
 ratio $\psi_{\gamma b}$ and those with constant mass fraction of
 free nucleons $\chi_{nuc}$  are plotted. The phase boundary between
 heavy nuclei and 
 uniform evaporated matter for the case $Y_e=0.3$ is described by the
 dot-dashed line with the
 label $\mu_{np,A}=\mu_{np,G}$.}
\label{fig1}
\end{figure}

We adopt a simplified, analytical equation of state for hot, dense matter given by 
\citet{LS}, in which the heavy nuclei are treated as being in a b.c.c. lattice. In 
accordance with the Wigner-Seitz approximation each heavy ion is considered to be 
surrounded by a charge-neutral spherical cell consisting of a less dense vapor of 
neutrons, protons and alpha particles as well as electrons. The volume of this cell 
is given by $V_c=n_A^{-1}$, where $n_A$ is the number density of heavy nuclei. We 
construct a three-dimensional table of the relevant thermodynamic quantities, i.e., 
$n_n, n_p, n_\alpha, \mu_n - \mu_p, A, Z, P_b$, as a function of the inputs ($\rho, 
T, Y_e$), where $n_n, n_p, n_\alpha$ are the number densities of free neutron, free 
proton, $\alpha$ particle, $\mu_n - \mu_p$ is the difference of chemical potential 
between the neutron and the proton, $A$ and $Z$ are the mean mass number and proton 
number of heavy nuclei. The contours of chemical potential, $\mu_n -
\mu_p$, are depicted in Fig.9.

\begin{table*}
\begin{minipage}{180mm}
\caption{Netrino reactions}
\label{tbl-1}
\begin{tabular}{@{}llll}
No. & Neutrino reaction & Reaction process & References \\
\hline
1 &$e + p \rightleftharpoons \nu_e + n$&$\beta$ process  & Burrows(2001)  \\
2 &$e^+ + n \rightleftharpoons \bar{\nu}_e + p$&positron capture on neutron & Burrows(2001)  
\\
3 &$e +A \rightleftharpoons \nu_e+A' $&electron capture on nuclei & Bruenn(1985) \\
4 &$\nu_i+N  \to \nu_i+N $&neutrino-nucleon scattering & Burrows(2001)  \\
5 &$\nu_i+A \to \nu_i+A $&neutrino-nucleus scattering  & Bruenn(1985) \\
6 &$\nu_i+e \to \nu_i+e $&neutrino electron scattering  & Burrows(2001)  \\
7 &$e+e^+\rightleftharpoons \nu_i+\bar{\nu}_i$&electron pair process & 
Burrows(2001) \\ 
8 &$N+N \rightleftharpoons N+N+\nu_i+\bar{\nu}_i$&nucleon-nucleon 
bremstrahlung&Raffelt(2001) \\
9 &$n+n \to n+p+e+\nu_e$ &Urca process & Shapiro and Teukolsky(1983) \\
10 &$\gamma \to \nu_i + \bar\nu_i$ &plasma decay &Ruffert et al.(1996)\\
\hline
\end{tabular}
\end{minipage}
\end{table*}

\subsection{Reactions of neutrinos}
The thermal and dynamical properties of a massive accretion disk are greatly 
influenced by neutrinos. The neutrino loss cools the disk and the interaction with 
ambient matter may produce the convection. In this paper the dominant neutrino reactions in the stellar core collapse are included, which are listed in Table 1. The summaries of the
various neutrino cross sections of relevance in supernova theory are
given in \citet{TS}, \citet{B85} and \citet{B01}.   
The approximate blockings in the lepton phase space 
are described in \citet{RJS}. Here the leptons
spectra are assumed to be represented by thermally equilibrium
Fermi-Dirac distributions with the temperature, $T$ (taken equal to the
gas temperature), and the degeneracy factor, $\eta_{l}$. The neutrino cross
sections, opacities, emissivities and reaction rates used in our model
are listed in Table A1--5(Appendix A). When the 
density $n_l$ and temperature $T$ of leptons are given, their chemical potential $\mu_l$ is determined
through the Fermi
integral, $n_l(\eta_l)=4\pi(kT/hc)^3 \int_0^\infty dx
x^2/[1+exp(x-\eta_l)] $. The electron-positron pair is assumed to be
in equilibrium with thermal photons.

The reaction and transfer of neutrinos
are evaluated in the simplified disk with the thickness $H(r)$ in which the
matter is assumed to be homogeneously distributed in the vertical 
direction. The thickness $H(r)$ is taken approximately to be a scale
height. The change of lepton fraction, $\dot{Y}_l$, is calculated along
the accreting flow. At the inner region of the disk, the hot, dense
matter changes the lepton fraction within the duration time of the
matter moving along the distance $\Delta r$ with the accreting velocity
$v_{accrt}$, $\Delta t=\Delta r/v_{accrt}$. When $v_{accrt} \approx
10^{-3}c \approx 10^{7}$[cm/sec] and the section of the moving distance is taken to be $\Delta
r \approx 0.1r_{ms} \approx 10^5(M_{BH}/3{\rm M}_\odot)$[cm], the
duration time is $\Delta t \approx 10$[ms]. In this time 
$\Delta t$ the neutrino reactions do not always reach to the local
thermodynamic equilibrium(LTE). The changes of leptons are culculated
without LTE.   
  
The changing rate of each lepton number in unit volume, $\dot{n}_l$, is produced 
through the networks of reactions,
\begin{eqnarray}
\dot{n}_e &=& -\dot{n}_\beta +\dot{n}_{\bar\beta}-\dot{n}_A +\dot{n}_{URCA}^+ ,\\
\dot{n}_{e^+} &=& -\dot{n}_{\bar\beta} -\dot{n}_{e e},\\
\dot{n}_{\nu_e} &=& \dot{n}_\beta + \dot{n}_A + \dot{n}_{ee} +\dot{n}_{NN} 
+\dot{n}_{URCA}^+ + \dot{n}_\gamma^+,  \\
\dot{n}_{\bar{\nu}_e} &=& \dot{n}_{\bar\beta} + \dot{n}_{ee} +\dot{n}_{NN} + 
\dot{n}_\gamma^+,  \\
\dot{n}_{\bar{\nu}_x} &=& \dot{n}_{ee} +\dot{n}_{NN} + \dot{n}_\gamma^+. 
\end{eqnarray}
Here $\dot{n}_\beta, \dot{n}_{\bar\beta}, \dot{n}_{ee}, \dot{n}_A$ and
$\dot{n}_{NN}$ are the net production rates of neutrinos by the
following reactions; 
the $\beta$ process, $e + p \rightleftharpoons \nu_e + n$, the positron
capture on neutron, $e^+ + n \rightleftharpoons \bar{\nu}_e + p$, the
electron pair annihilation, $e+e^+\rightleftharpoons \nu_i+\bar{\nu}_i$, the
electron capture on the nuclei, $e +A \rightleftharpoons \nu_e+A'$ and
the nucleon-nucleon bremstrahlung, $N+N \rightleftharpoons
N+N+\nu_i+\bar{\nu}_i$. The positive sign in these net rates indicates the
increase of neutrinos. The plus sign at the upper suffix denotes the
production rates of neutrinos without inverse reaction, i.e., $\dot{n}_{URCA}^+$ and
$\dot{n}_\gamma^+$ are the production rates by the
URCA process, $N+N \to n+p+e+\nu_e$, and by the plasma decay, $\gamma
\to \nu_i + \bar\nu_i$.  

The changes of neutrino density are also brought about by the leakage of
 neutrinos from the disk. The flux number density of each neutrino at
 the surface of the 
disk, $F_{n(\nu_i)}$, are approximately expressed by the production rate of 
neutrino, $\dot{n}_{\nu_i}^+$, and the absorptive, scattering and total opacities,
$\kappa_{\nu_i a}, \kappa_{\nu_i s}$ and $\kappa_{\nu_i}=\kappa_{\nu_i
a}+\kappa_{\nu_i s}$,   
\begin{equation}
F_{n(\nu_i)}=\frac{\dot{n}^+_{\nu_i}H}{1.5\tau_{\nu_i 
a}\tau_{\nu_i}+\sqrt{3}\tau_{\nu_i a} + 1},   
\end{equation}
where $\tau_{\nu_i}$ and $\tau_{\nu_i a}$ are the total and absorptive 
optical depths for $\nu_i$ (see Appendix B). 
The absorptive and scattering opacities for each type of neutrino are
given by
\begin{eqnarray}
\kappa_{\nu_e a} &=& \kappa^a_{\beta} + \kappa^a_{A} + \kappa^a_{NN}, \quad 
\kappa_{\bar{\nu}_e a} =\kappa^a_{\bar\beta}+ \kappa^a_{NN}, \\
\kappa_{{\nu}_x a} &=& \kappa^a_{NN}, \quad 
\kappa_{\nu_i s} = \kappa_n^s + \kappa_p^s + \kappa_A^s + \kappa_e^s,
\end{eqnarray}
where $\kappa_n^s, \kappa_p^s, \kappa_A^s$ and $\kappa_e^s$ are the scattering 
opacities due to the scattering with neutron, proton, nucleus and
electron which are represented in Table A2.

Thus the change of the each neutrino fraction, $\dot{Y}_{\nu_i}$, along the 
accreting flow is expressed by
\begin{eqnarray}
n_b\dot{Y}_{\nu_i}=v^r\nabla_r n_{\nu_i}= \dot{n}_{\nu_i} - 
\frac{F_{n(\nu_i)}}{H}.
\end{eqnarray}
The changes of the electron fraction, $\dot{Y}_{e}$, and of the
positron fraction, $\dot{Y}_{e^+}$, are given by
\begin{eqnarray}
n_b\dot{Y}_{e} = \frac{\dot{n}_e}{n_b}, \quad \dot{Y}_{e^+} = \frac{\dot{n}_{e^+}}{n_b} + v^r\nabla_r 
\frac{n_{e^+th}}{n_b}, 
\end{eqnarray}
where $n_{e^+th}$ is the number density of positrons which
is equilibrium with thermal photons(see Appendix C).

The energy loss carried out by neutrinos is similarly expressed with 
the emissivities of neutrinos, $q_{\nu_i}$. The emissivity of each type of 
neutrino is described as
\begin{eqnarray}
q_{\nu_e} &=& q_\beta + q_{ee} + q_A + q_{NN} + q_{URCA} + q_\gamma, \\ 
q_{\bar{\nu}_e} &=& q_{\bar\beta} + q_{ee} + q_{NN} + q_\gamma, \\
q_{\nu_x} &=& q_{ee}(\nu_x) + q_{NN}(\nu_x) + q_{\gamma}(\nu_x),
\end{eqnarray}
where $q_\beta, q_{\bar\beta}, q_{ee}, q_A, q_{NN}, q_{URCA}$ and $q_\gamma$ are
emissivities by the $\beta$ process, the positron capture on neutron,
the electron pair annihiration, the electron capture on the nuclei, the
nucleon-nucleon bremstrahlung, the URCA process, and by the plasma
decay; $q_{ee}(\nu_x), q_{NN}(\nu_x), q_{\gamma}(\nu_x)$ are those for
heavy neutrinos, $\nu_x = \nu_\mu, \nu_\tau$ (see Table A3).    
The energy flux density of each neutrino at the surface of the disk, 
$F_{\varepsilon(\nu_i)}$, is then given by
\begin{equation}
F_{\varepsilon(\nu_i)}=\frac{q_{\nu_i}H}{1.5\tau_{\nu_i a}\tau_{\nu_i} + 
\sqrt{3}\tau_{\nu_i a}+1}.
\end{equation}

\section{Relativistic Model of Accretion Disk with Neutrino Loss}
The duration time of GRBs requires that the
collapsed, accreting matter has large specific angular momentum. The
central black hole may be rapidly rotating\citep{MW}. We
derive precisely the shear stress tensor $\sigma_{\varphi r}$ in Kerr
spacetime. The relativistic model for the accretion disk was
investigated by \citet{NT73}. However, their model is
insufficient in some respects. The boundary condition at the inner disk
is not suitable, i.e.,
the thickness of the disk becomes infinitesimal and the density diverges
to infinity at the boundary. The work produced by
advection was not included in their model. We reconstruct the
relativistic model without singularity and with 
advectional motion based on the standard theory
of the thin accretion disk. 

We choose units with $G = c = 1$. The metric of spacetime, in terms of Boyer-Lindquist 
time $t$ and any arbitrary spatial coordinates $x^j$, has the form
\begin{equation}
ds^2 = -\alpha^2dt^2 + g_{jk}(dx^j+\beta^jdt)(dx^k+\beta^kdt).
\end{equation}
The metric coefficients $\alpha, \beta^j,g_{jk}$ are the lapse, shift, and 3-metric 
functions. Theses functions are given by
\begin{eqnarray}
\alpha &=& \Bigl(\frac{\Sigma\Delta}{A}\Bigr)^{1/2}, \quad 
g_{rr}=\frac{\Sigma}{\Delta}, \quad g_{\theta\theta}=\Sigma, \quad 
g_{\varphi\varphi}=\frac{\sin^2{\theta}A}{\Sigma} \nonumber \\  
\beta^\varphi &=&  -\omega = -\frac{2Mar}{A}, \quad \beta^j=g_{jk}=0 \quad {\rm 
for\ all\ other}\ j{\rm\ and}\ k, \nonumber
\end{eqnarray}
where $M$ is the mass of the black hole, $a$ is its angular momentum per unit mass, 
and the functions $\Delta,\Sigma,A$ are defined by
\begin{eqnarray}
\Delta = r^2 - 2Mr + a^2, \quad \Sigma=r^2+a^2\cos^2{\theta}, \\ 
A=(r^2+a^2)^2-a^2\Delta\sin^2{\theta}.  \nonumber
\end{eqnarray}

We introduce a set of local observers who rotate with a Keplerian,
circular orbit. Each observer carries an orthonormal tetrad. For the
Keplerian orbiting observer with a coordinate angular velocity of $\Omega$,
its world line is $r=$constant, $\theta=$constant, $\varphi=\Omega t$+
constant. The observer in the LNRF who rotates with the angular velocity $\omega$
measures the linear velocity of a particle moving in a Keplerian
orbit as $v_{(\varphi)}=\alpha^{-1} \sqrt{g_{\varphi\varphi}}(\Omega
- \omega)$. Its Lorentz factor is ${\gamma}=(1-
v_{(\varphi)}^2)^{-1/2}$. We adopt the Keplerian orbiting, observer's
frame (``orbiting frame'') with the set of it's basis vectors:
\begin{eqnarray}
{\bf e}_{\hat 0} &=& \gamma e^{-\nu}\big(\frac{\partial}{\partial 
t}+\Omega\frac{\partial}{\partial \varphi}\big), \\
{\bf e}_{\hat\varphi} &=& \gamma \Bigl(e^{-\psi} 
\frac{\partial}{\partial 
\varphi}+ v_{(\varphi)}e^{-\nu}(\frac{\partial}{\partial 
t}+\omega\frac{\partial}{\partial \varphi})\Bigr), \\ 
{\bf e}_{\hat r} &=& e^{-\mu_1}\frac{\partial}{\partial r}, \quad
{\bf e}_{\hat \theta}= e^{-\mu_2}\frac{\partial}{\partial 
\theta},
\end{eqnarray}
where $e^\nu, e^\psi, e^{\mu_1}, e^{\mu_2}$ are difined by $e^\nu=\alpha, 
e^{2\psi}=g_{\varphi\varphi}, e^{2\mu_1}=g_{rr}, e^{2\mu_2}=g_{\theta\theta}$.

The corresponding orthogonal basis of one-forms is
\begin{eqnarray}
&{\vec \omega}^{\hat{t}} = \hat{\alpha}{\rm\bf d}t 
- v_{(\varphi)}\gamma e^\psi {\rm\bf d}\varphi, \quad
{\vec \omega}^{\hat{\varphi}} = -\Omega \gamma e^\psi {\rm\bf d}t 
+  \gamma e^\psi
{\rm\bf d}\varphi \nonumber \\
&{\vec \omega}^{\hat{r}} = e^{\mu_1}{\rm\bf d}r, \quad 
{\vec \omega}^{\hat{\theta}} = e^{\mu_2}{\rm\bf d}\theta,
\label{eq: one-form}
\end{eqnarray}
where $\hat{\alpha}$ is the ``lapse function'' in an orbital frame,
$\hat{\alpha}=\alpha\gamma(1+ v_\omega v_{(\varphi)})$.

The shear stress in an orbital frame is expressed by
\begin{equation}
\sigma^{\hat{r}\hat\varphi} = e_\alpha^{\hat{r}}e_\beta^{\hat{\varphi}} 
\sigma^{\alpha\beta},
\end{equation}
where $\sigma^{\alpha\beta}$ is the shear stresses in the Boyer-Lindquist coordinate, 
and $e_\alpha^{\hat{a}}$ is the Boyer-Lindquist components of the basis of one-forms 
$(\ref{eq: one-form})$ in orbital frame, ${\vec \omega}^{\hat{a}} = 
e^{\hat{a}}_\alpha {\rm\bf d}x^\alpha$. The shear stress in the Boyer-Lindquist 
coordinate is defined by
\begin{equation}
\sigma_{\alpha\beta} \equiv \frac{1}{2}(u_{\alpha; \mu} h^{\mu}_{\beta} + u_{\beta; 
\mu} h^{\mu}_\alpha ),
\end{equation}
where $h_{\mu\nu}$ is the projection tensor, $h_{\mu\nu}=g_{\mu\nu}+u_\mu u_\nu$.
The shear stress in the orbiting frame is then represented as
\begin{eqnarray}
\sigma^{\hat{r} \hat\varphi} = \frac{1}{2} \Bigl\{ \frac{1}{\sqrt{g_{rr}}} 
( \gamma^2 v_{(\varphi)} \frac{\partial \ln (\Omega-\omega)}{\partial r} + 
\frac{1}{2} v_\omega \frac{\partial \ln \omega}{\partial r} )  \Bigr\}, 
\end{eqnarray}
where $v_\omega$ is the rotating velocity of spacetime,
$v_\omega = \alpha^{-1} \sqrt{g_{\varphi\varphi}}\omega$. 
While the absolute value of the shear stress in the Shwarzchield
spacetime monotonously increases as 
$|\sigma^{\hat{r} \hat\varphi}|\approx 0.75\Omega$ according to $r\to
r_{ms}$,  the shear stress in rapidly rotating spacetime has the minimum in the vicinity of $r_{ms}$. 

The generation rate of the viscous heating is evaluated in the orbiting frame by
\begin{equation}
\dot\epsilon =-t_{\hat{i}\hat{j}}\sigma^{\hat{i}\hat{j}} ,
\end{equation}
where $t_{\hat{i}\hat{j}}$ is the stress tensor. For the ``$\alpha$ model'' of the 
viscosity the stress tensor is expressed by 
$t_{\hat{i}\hat{\varphi}}=\alpha_{vis} P_{th}$, where $P_{th}$ is the
thermal pressure and 
$\alpha_{vis}$ is a parameter with $\alpha_{vis}=0.1-0.01$.

Then we consider the energy equation in the orbiting frame. The
energy-momentum tensor of the viscous fluid with heat flux is
expressed\citep{MTW,NT73,Y95} as follows,
\begin{eqnarray}
T^{\hat\alpha\hat\beta} =(\rho+\epsilon)u^{\hat\alpha}u^{\hat\beta} + 
Ph^{\hat\alpha\hat\beta} + t^{\hat\alpha\hat\beta} + q^{\hat\alpha}u^{\hat\beta} + 
u^{\hat\alpha}q^{\hat\beta},
\end{eqnarray}
where  $\rho, \epsilon$, and $P$ are the rest-mass density, internal 
energy density and pressure and $q^{\hat\alpha}$ is the heat
 flux carried by neutrinos and photons. 
By using the covariant derivative, $\vec\nabla$, 
the local energy conservation, $\vec{u}\cdot (\vec\nabla\cdot{\rm\bf T})=0$, is 
represented as
\begin{eqnarray} 
\frac{\partial \epsilon u^{\hat 0} + q^{\hat 0}}{\partial x^{\hat{0}}} +  
\frac{1}{\hat\alpha} \nabla_{\hat{i}} \cdot \hat\alpha (\epsilon { u}^{\hat{i}} +  
{ q}^{\hat{i}}) + P\frac{\partial u^{\hat 0}}{\partial x^{\hat 0}} \\  + 
\frac{P}{\hat\alpha} \nabla_{\hat{i}} \cdot \hat\alpha{ u}^{\hat{i}}  + 
\sigma^{\hat\alpha \hat\beta}t_{\hat\alpha \hat\beta} = 0.
\end{eqnarray}

The mass conservation $\vec\nabla\cdot(\rho\vec{u})=0$ is written by 
\begin{eqnarray} 
\frac{\partial \rho u^{\hat 0}}{\partial x^{\hat 0}} + \frac{1}{\hat\alpha} 
\nabla_{\hat{i}} \hat\alpha \rho { u}^{\hat{i}}  = 0.
\end{eqnarray}
By using the projection vector in $\varphi$ direction, {\bf 
h}$^{\varphi}=h^{\varphi \beta}$, the Euler equation in the 
$\varphi$ direction, ${\rm\bf h}^{\varphi}\cdot (\nabla\cdot{\rm\bf 
T})=0$, is expressed in the frame fixed at the distant stars,
\begin{eqnarray} 
\frac{\partial (\rho+\epsilon+p) u^{0}u_{\varphi}}{\partial t} + \frac{1}{\alpha} 
\nabla_{i} \big(\alpha (\rho+\epsilon+ P){u}^{i}u_{\varphi}+ t^i_\varphi + 
u_{\varphi}{ q}^{i} \big)  \\ + \frac{\partial\alpha p}{\alpha\partial x^{\varphi}} = 
0. \label{eq:angular}
\end{eqnarray}
  
We are interesting in the gross properties of the disk with neutrino loss. For 
simplicity, we consider a steady state disk with axially symmetry.  As in the standard 
theory of thin accretion disks, we consider height-averaged quantities. Putting the 
integrated stress as $W= \int_{-h}^ht_{\hat{r}\hat\varphi}dz=\int_{-h}^h 
\alpha_{vis}P_{th}dz$, and introducing the flux density at the surface of the disk, 
$F=q^{\hat z}(z= H)$, the energy equation is expressed by
\begin{eqnarray}
F+\Bigl(\sigma_{{\hat r}{\hat\varphi}} + \frac{1}{\hat\alpha\alpha_{vis}} 
(\nabla_{\hat{r}} \hat\alpha u^{\hat{r}} + P^{-1}
\nabla_{\hat{r}} \hat\alpha\epsilon u^{\hat{r}})\Bigr)W =0. \label{eq:energy}
\end{eqnarray}

From the equation $(\ref{eq:angular})$ the conservation of angular momentum is 
\begin{equation}
\nabla_{i} \cdot \big(\alpha (\rho+\epsilon+ P){u}^{i}u_{\varphi}+ t^i_\varphi + 
u_{\varphi}{q}^{i} \big)=0
\end{equation}
The angular momentum is transported by the viscous stress $t^i_\varphi$ and by the 
neutrino flowing $q^i$. Introducing the total mass flux $\dot{M}$, we express the 
transport equation of angular momentum as
\begin{equation}
\frac{\partial}{\partial r}\big( -\frac{\dot M}{2\pi}u_\varphi + \sqrt{\Delta} 
\gamma e^\psi W) - 2ru_\varphi F =0.
\end{equation}
It has the solution:
\begin{equation}
W= \frac{\dot{M}}{2\pi}\Omega\Pi, \label{eq:angular-moment}
\end{equation}
where $\Pi$ is the function with the value of unity at the great distance from the 
hole:
\begin{eqnarray}
\Pi = \frac{{u}_\varphi+ C}{\sqrt{{\Delta}} {\gamma} e^{\psi}{\Omega}} + 
\frac{2\pi fg}{{\Omega}\dot{M}}. 
\end{eqnarray}
The term $fg$ is due to the neutrino flowing. The functions $f$ and $g$ are  
\begin{eqnarray}
f&=& -\frac{\dot{M}}{2\pi}\int^\infty_r \frac{2\varpi {u}_\varphi 
\tilde{\sigma}_{\hat{r}\hat{\varpi}}} {g{\Delta}{{\gamma}}^2 e^{2\psi}} 
({u}_\varphi + C) d\varpi, \\
g&=& \frac{1}{\sqrt{\Delta}\gamma e^{\psi}}\exp\biggl[ -\int^\infty_r 
\frac{2\varpi {u}_\varphi \tilde{\sigma}_{\hat{r}\hat{\varphi}}}{\sqrt{{\Delta}} 
{\gamma} e^{\psi}} d\varpi\biggr], \\
\tilde{\sigma}_{\hat{r}\hat{\varphi}} &=& {\sigma}_{\hat{r}\hat{\varphi}} \Bigl(1 + 
(1+ \frac{\epsilon}{P}) \frac{\sqrt{\Delta} u^{\hat r}} {2 
\sigma_{\hat{r}\hat{\varphi}} \alpha_{vis}\varpi^2} (\frac{\partial \ln{\sqrt{\Delta} 
u^{\hat r}}} {\partial \varpi} \\ &+& \frac{1}{1+ P/\epsilon} \frac{\partial \ln\epsilon} 
{\partial \varpi})\Bigr).
\end{eqnarray}
Here $C$ is the integral constant. Novikov $\&$ Thorne(1973) adopted such as
$C=-{u}_\varphi(r_{ms})$ since they assumed no viscous stress can act at $r=r_{ms}$. 
The accreting matter with viscosity continuously flows into a black
hole through the boundary radius  
$r_{ms}$\citep{Y95}. We set the constant $C$ so that the accreting
matter continuously flows through the radius $r_{ms}$ and then the viscous stress has the value 
such as $t_{\hat{r}\hat\varphi}(r_{ms}) = \alpha_{vis}p(r_{ms})$, 
\begin{eqnarray}
C= - {u}_\varphi(r_{ms}) 
+ \frac{4\pi h}{\dot{M}}\sqrt{{\Delta}(r_{ms})} {\gamma}(r_{ms}) 
e^{\psi}(r_{ms})\alpha_{vis}{P}(r_{ms}).
\end{eqnarray}

Dynamically equilibrium structure of the disk is given by the
Euler equation. The equation of motion of the fluid, $h^k_iT^{j}_{k;j}=0$,
for the steady, axially symmetric and rotating fluid can be expressed in the
total differential form:
\begin{equation}
\frac{dp}{\rho+P}=\gamma^2(-d\nu+v_{(\varphi)}^2d\psi-v_{(\varphi)}v_\omega 
d\ln\omega) \equiv -dU. \label{eq dp}
\end{equation}
The equipressure surfaces are given by the equation $U=$constant. The
quantity $U=U(p)$ is equal in the Newtonian limit to the total potential
(gravitational plus centrifugal) expressed in the units of ${\rm c}^2$ \citep{AJS78}. 
In the neighborhood of equatorial plane the potential $U$ is expanded with $z$:
\begin{equation}
U = U_0(r,0) + \frac{1}{2}\Omega^2_0 z^2, 
\end{equation}
where $\Omega^2_0$ is 
\begin{equation}
\Omega^2_0 = \frac{{\rm G}M}{r^3}\Phi(r).
\end{equation}
The function $\Phi(r)$ becomes unity far a way from a hole. Its explicit 
expression of $\Phi(r)$ is shown in Appendix D.   
The thickness of the disk is approximately given by the scale height,
\begin{eqnarray}
H&=&\sqrt{2\frac{P}{\rho}}\Omega_0^{-1}.
\end{eqnarray}

The energy equation ($\ref{eq:energy}$) integrated over the thickness of the disk 
is represented with the mass accretion rate, $\dot{M}= -4\pi r H \rho 
\hat\alpha u^{\hat r}$,
\begin{eqnarray}
F + \sigma_{{\hat r}{\hat\varphi}} W + \frac{\dot M}{4\pi 
r^2} \Bigl( \frac{\partial}{\partial \ln{r}} \frac{\epsilon + P}{\rho} 
 - \frac{P}{\rho} \frac{\partial\ln P}{\partial \ln{r}}\Bigr) = 0 
\label{eq: Energ}
\end{eqnarray}
 
The flux density $F$ is evaluated by the simplified neutrino transfer,
\begin{equation}
F= \sum_{\nu_i}\frac{q_{\nu_i} H}{1.5\tau_{\nu_i a}\tau_{\nu_i} +
\sqrt{3}\tau_{\nu_i a} + 1}, \label{F}
\end{equation}
where $\nu_i$ means the flaver of neutrino, $\nu_i=\nu_e, \bar{\nu}_e,
\nu_\mu, \bar{\nu}_\mu,  
\nu_\tau, \bar{\nu}_\tau$. Thus the basic equations $(\ref{eq: Energ})$ and
$(\ref{eq:angular-moment})$ determine the stationary, thermally
equilibrium structure of accretion disk.  

In addition to the above formalism of accretion disk, we shall consider
the thermal energy of the partially degenerated matter which contributes to the
viscous heating and the advection. 
In the ranges of density and
temperature, $10^{11} < \rho < 10^{13}$(g/cm$^3$) and $10^{10} < T <
10^{11}$(K), the degeneracy factors of electrons has the
values in the range, 
$5<\eta_e < 15$ (see Fig. 10). The pressure and the internal energy of
electrons are expressed as
\begin{eqnarray}
P_e &=& \frac{\mu_e}{12\pi^2}\biggl(\frac{\mu_e}{\hbar
c}\biggr)^3\biggl[1 + \biggl( \frac{kT}{\mu_e}\biggr)^2 \biggl(2\pi^2
-3 \biggl(\frac{m_ec^2}{kT}\biggr)^2 \biggr)  \nonumber \\
&+& \biggl(\frac{kT}{\mu_e}\biggr)^4 \pi^2 \biggl(\frac{7}{15}\pi^2 - \frac{1}{2}\biggl( \frac{m_ec^2}{kT} \biggr)^2 \biggr) \biggr], \\
\epsilon_e &=& \frac{\mu_e}{4\pi^2} \biggl(\frac{\mu_e}{\hbar c}
\biggr)^3 \biggl[ 1+ \biggl(\frac{kT}{\mu_e} \biggr)^2 \biggl( 2\pi^2
- \biggl(\frac{m_ec^2}{kT} \biggr)^2 \biggr)  \nonumber \\
&+& \biggl(\frac{kT}{\mu_e} \biggr)^4 \biggl( \frac{7}{15}\pi^2 - \frac{1}{2} \biggl( \frac{m_ec^2}{kT} \biggr)^2 \biggr) \biggr].
\end{eqnarray}
The first terms in the both square brackets provide the pressure and 
 internal energy corresponding to the completely degenerated
state. The second and third terms depend on the temperature. We
 introduce the thermal pressure of electrons, $P_{e th}$, expressed by the above
expression without the first term. The completely degenerate matter
has no contribution in cooling or heating by advectional motion.
Thus the pressure and internal energy density without the
 completely degenerate parts are used for the advection or for the $\alpha$ model of viscosity. 

The baryonic matter is hardly degenerated over the above phase area in
 $\rho$ --- $T$
 plane. The pressure of baryonic matter is
consist of that of free nucleons $P_{nuc}$, that of $\alpha$ particles $P_\alpha$, that of heavy nuclei
\citep{LS},
\begin{equation}
P_b=P_{nuc}+P_\alpha -\beta(D-uD')+\frac{un_A}{A}h[kT(1-u)-\mu_A u],
\end{equation}
where $\mu_H$ is the chemical potential of heavy nuclei, $u$ is the fraction of space 
occupied by heavy nuclei, $\beta, D, D', h$ are the functions of nuclear quantities 
\citep{LS}. The values of $P_b$ in the above phase area are
almost same as $n_bkT$. Thus the internal energy of baryonic
matter is approximately set to be $\epsilon_b \approx 3/2P_b$. The total
pressure and internal energy density are given by
\begin{eqnarray}
P &=& P_b +P_e + aT^4(\frac{11}{12} + \frac{7}{8}\tau), \\
\epsilon &=& \epsilon_b + \epsilon_e +aT^4(\frac{11}{4} + \frac{21}{8}\tau),
\end{eqnarray}
where $a$ is the radiation constant $\tau$ is the neutrino opacity,
$\tau=\Sigma(\tau_{\nu_i}/2 + 1/\sqrt{3})/ (\tau_{\nu_i}/2 + 1/\sqrt{3}
+1/3\tau_{\nu_i a})$. The third terms on the right hand side represent the
contributions of radiation in thermally equilibrium with relativistic electron positron pairs and of neutrinos\citep{PN95}

\section{RESULTS}
We have solved the stationary flow structure at the position range, $r=r_{ms} \sim
10^2r_{ms}$. At the outer boundary of the calculation we select the most
large value of $Y_e$ within $Y_e\le 0.5$ which satisfies the stationary
conditions, (\ref{eq: Energ}) and (\ref{eq:angular-moment}). The
fractions of neutrinos at the boundary, $Y_{\nu_i}$, are taken which
provide the local thermodynamical equilibrium in neutrino reactions. The
accretion rate $\dot M$ is taken to be $\dot  
M = 0.01 \sim 1 {\rm M}_\odot$sec$^{-1}$, and the angular momentum
parameter of a black hole ranges from $a=0 \sim 1$. The accreting
velocity, $v\equiv u^r/u^0$, is derived from the mass conservation,
$\dot M = const.$, for given density $\rho$ and height $H$. The
typical parameters in the models of GRBs are selected as $a=0.9, \dot
M = 0.1{\rm M}_\odot$sec$^{-1}$, and $M_{BH}=3{\rm
M}_\odot$\citep{MW}. The viscous parameter is set to be
$\alpha_{vis}=0.05$\citep{T04}.

\begin{figure}
\includegraphics[width=8cm,height=6cm]{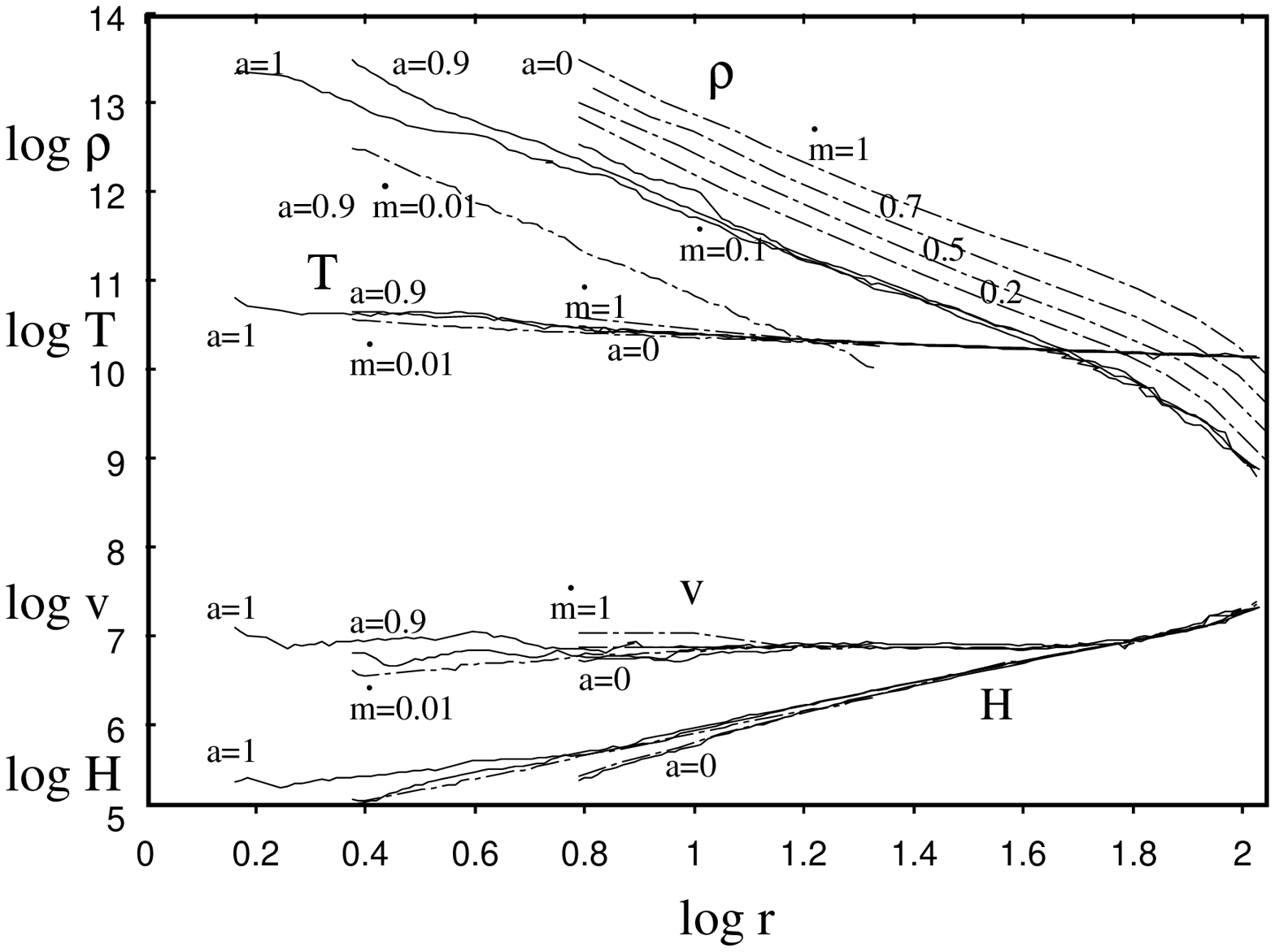}
\caption{The flow profiles of density $\rho$(g cm$^{-3}$), temperature $T$(K), drift velocity
 $v$(cm/sec), scale hight $H$(cm). The radius $r$ is normalized by the gravitational
 radius $r_g$. The mass of a black hole is fixed to be $M_{BH}=3{\rm M}_\odot$,
 and the angular momentum parameters are selected such as $a=0,0.9,1$. The numbers denote the accretion rates,
 $\dot M$= 1,0.5,0.2,0.1,0.01(${\rm M}_\odot$  sec$^{-1}$). The flow profiles with
 $\dot{M}=0.1{\rm M}_\odot$sec$^{-1}$ are plotted by filled lines, and those
 with $\dot{M}=1{\rm M}_\odot$sec$^{-1}$ and  $\dot{M}=10^{-2}{\rm M}_\odot$sec$^{-1}$ are depicted by dot-dashed lines.
\label{fig2}}
\end{figure}

\begin{figure}
\includegraphics[width=7cm,height=5cm]{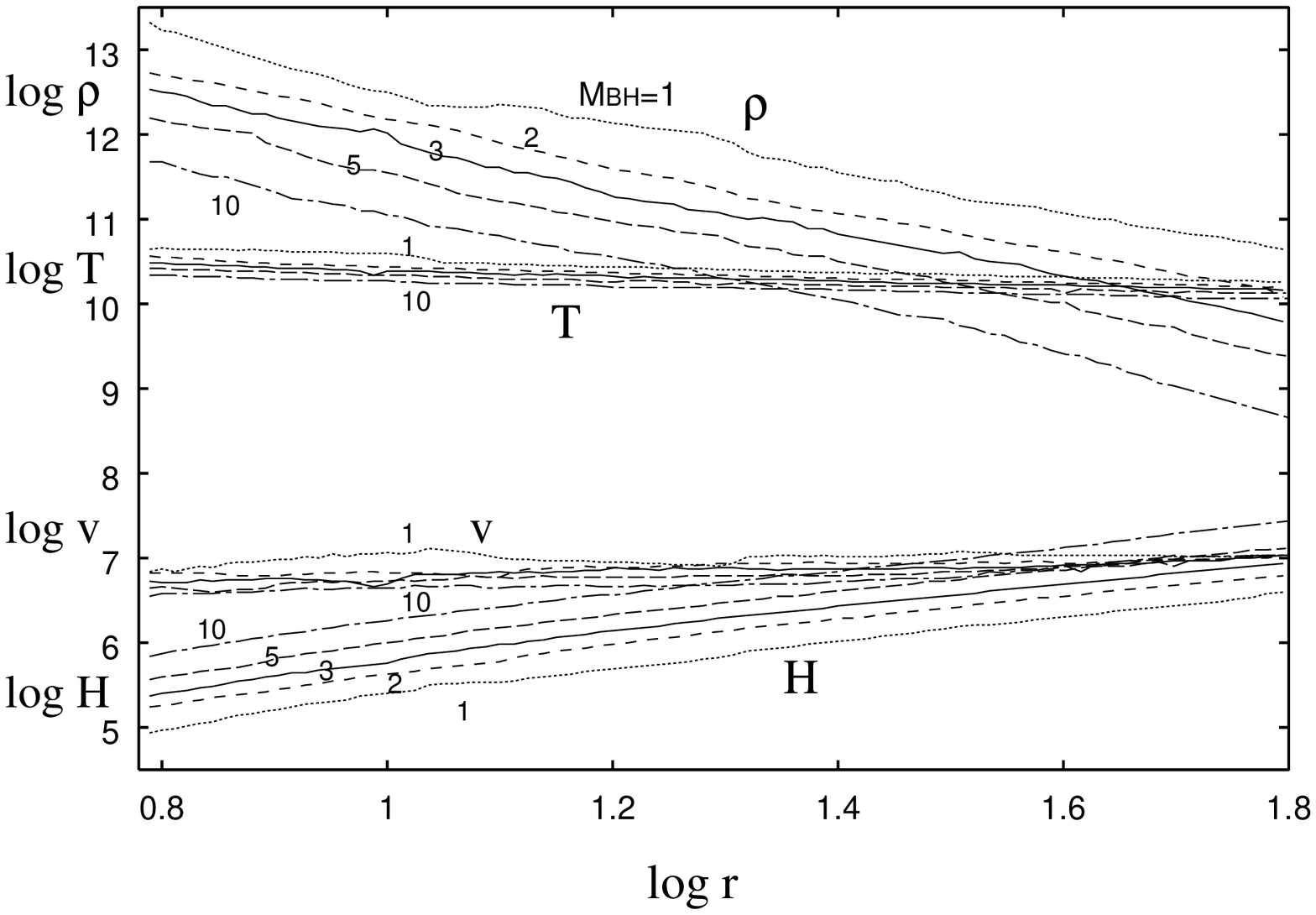}
\caption{The same as Fig. 2, but for the mass dependence of a black hole. The numbers denote the masses of a black hole,
 $M_{BH}$= 1,2,3,5,10(${\rm M}_\odot$). The accretion rate and the specific anguler momentum of a
black hole are fixed as
$\dot{M}=0.1{\rm M}_\odot$sec$^{-1}$ and $a$=0. The
 radius $r$ is normalized by the gravitational radius $r_g$. 
\label{fig3}}
\end{figure}

\subsection{Flow structures}
The flow profiles are shown in Fig. 2, where the density $\rho$, temperature 
$T$, accreting velocity $v$ and scale height $H$ are plotted in the
radius normalized by the gravitational radius, ${r_*}=r/r_g$. While the
density $\rho$ increases according to the accretion rate $\dot M$ and
reaches to a very high value, $\rho \ge 10^{13}$[g/cm$^3$] for $\dot{M}
\ge 0.1{\rm M}_\odot$sec$^{-1}$, the temperature, scale height and accreting
velocity are little changed for the wide range of accretion rate,
$\dot{M}=10^{-2} \sim 1{\rm M}_\odot$sec$^{-1}$. The profiles of temperature
and accreting velocity are nearly constant along the flow, $T \approx
3\times 10^{10}$K and $v \approx 10^{-3}c$. The scale height is
expressed as $H \approx 0.1 r_{g}(r_*)^{5/4}$, showing the very thin
disk formed around a black hole. 

These flow profiles are somewhat different from the previous works on
the advection dominant flow with neutrino loss\citep{PWF,NPK,MPN}. In
the previous cases the temperature increases in
proportion to the accretion rate. The profiles of the temperature and
velocity rapidly increase to be 
$T \ge 10^{11}$K and $v \to c$ for $r \to r_{ms}$. Then the density does not reach to
the extremly high value which is restricted to be $\rho \approx 10^{10
\sim 12}[{\rm g/cm}^3]$ in the ranges of accretion rate, $\dot{m}=0.1
\sim 10M_\odot$/sec. These differences 
in our results and previous ones are caused mainly by the cooling
process, which will be discussed later. In the previous 
cases the most of neutrinos are absorbed by free nucleons and thus
trapped in the accreting flow. In our study the antielectron neutrinos
$\bar{\nu}_e$ are little absorbed by free protons since free protons are
reduced in the fraction, $Y_p \le 10^{-3}$. The accretion disk is then
efficiently cooled by $\bar{\nu}_e$. It suppresses the rise of
temperature and velocity, which then produces the highly dense flows.  

The density $\rho$ and the temperatute $T$ decrease as the mass of a central black
hole increases when the accretion 
rate is restricted to be constant. The flow profiles are shown in
Fig. 3 where the mass $M_{BH}$ is 
changed from $1{\rm M}_\odot$ to $10{\rm M}_\odot$ and other parameters are fixed
to be $a=0,\quad \dot{M}=0.1{\rm M}_\odot$sec$^{-1}$. Their scale dependences
are described approximately as $T\propto M_{BH}^{-1/3}, \ H \propto
M_{BH}$ and $\rho \propto \dot{M}M_{BH}^{-2}$. That of the surface density
 is then $\Sigma=\rho H \propto M_{BH}^{-1}\dot{M}$. 
Thus the obtained flow profiles are expressed approximately by $\rho \approx
2\times 10^{14}r_*^{-2.5}\dot{m}m^{-2}$[g/cm$^3$] and $T \approx 5\times
10^{10}r_*^{-1/4}m^{-1/3}$K, where $m$ is the mass of a central black
hole normalized by $1{\rm M}_\odot$ and $\dot{m}$ is the accretion rate
normalized by $1{\rm M}_\odot$/sec.

\begin{figure}
\includegraphics[width=7cm,height=5cm]{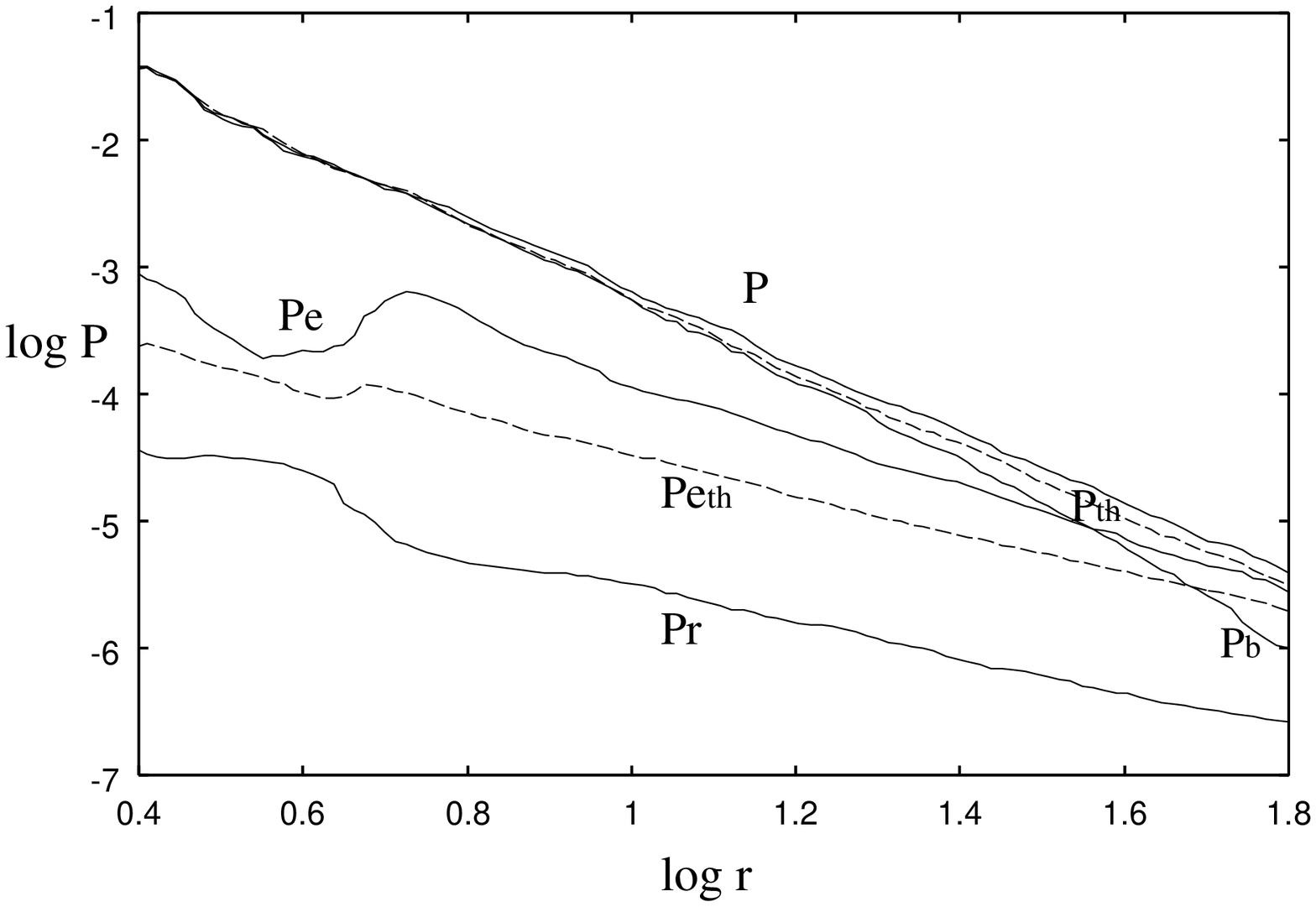}
\caption{The profiles of total pressure $P$, electron pressure $P_e$,
 pressure of baryonic matter $P_b$, radiation pressure $P_r$, total thermal
 pressure $P_{th}$, and the thermal pressure of electrons $P_{eth}$ are
plotted in the case of the accretion with $\dot{M}=0.1{\rm M}_\odot$ sec$^{-1}$, 
 $M_{BH}=3{\rm M}_\odot$ and $a=0.9$. The unit of
pressure $P$ is (MeVfm$^{-3}$).
\label{fig4}}
\end{figure}

The inner disk is supported mainly by the baryon pressure $P_b$. The
compositions of the pressure, $P_b, P_e, P_r$ and the thermal pressure,
$P_{th}, P_{e, th}$ are shown in Fig. 4. While the electron pressure
$P_e$ dominates at the outer region, $r \ge 15r_{ms}$, it more gradually increases than
$P_b$ according to the accreting flow. When $Y_e$ rapidly decreases, the
electron pressure itself decreases along the flow. The contribution of
the radiation pressure to the total pressure is negligibly small at the
inner region, but at the outer region, $r \ge 10^2r_g$, it becomes
comparable with that of $P_b$. The thermal pressure of electrons $P_{e
th}$ is comparable with its degenerate part of the pressure. The
degeneracy factor of electron $\eta_e$ is ranged as $\eta_e \approx 2
\sim 10$ for $\dot{m}=0.1$ which doesn't reach to an extremely large
value(see Fig. 10).

\begin{figure}
\includegraphics[width=7cm,height=5cm]{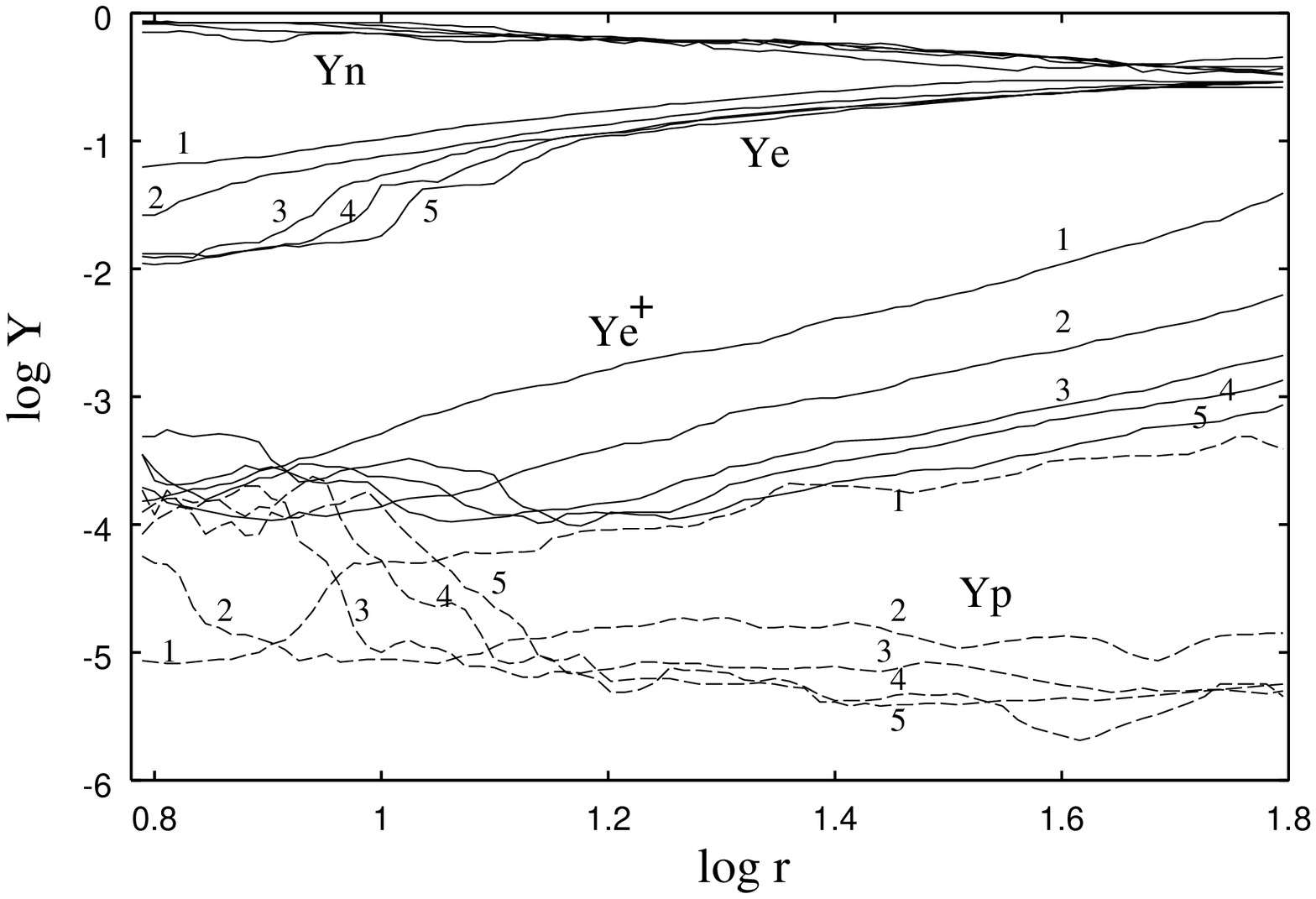}
\caption{The profiles of electron fraction $Y_e$, positron fraction
 $Y_{e^+}$, free neutron fraction $Y_n$, and free proton fraction
 $Y_p$ for the accretion onto a non-rotating black hole, 
$a=0$, with
 mass $M_{BH}=3{\rm M}_\odot$. The numbers of the label 1,2,3,4,5 denote the
 cases for accretion rates, $\dot{M}$ = 0.05,0.2,0.5,0.7,1($ {\rm M}_\odot$ sec$^{-1}$). The
 radius $r$ is normalized by the gravitational radius $r_g$.  
\label{fig5}}
\end{figure}

\begin{figure}
\includegraphics[width=7cm,height=5cm]{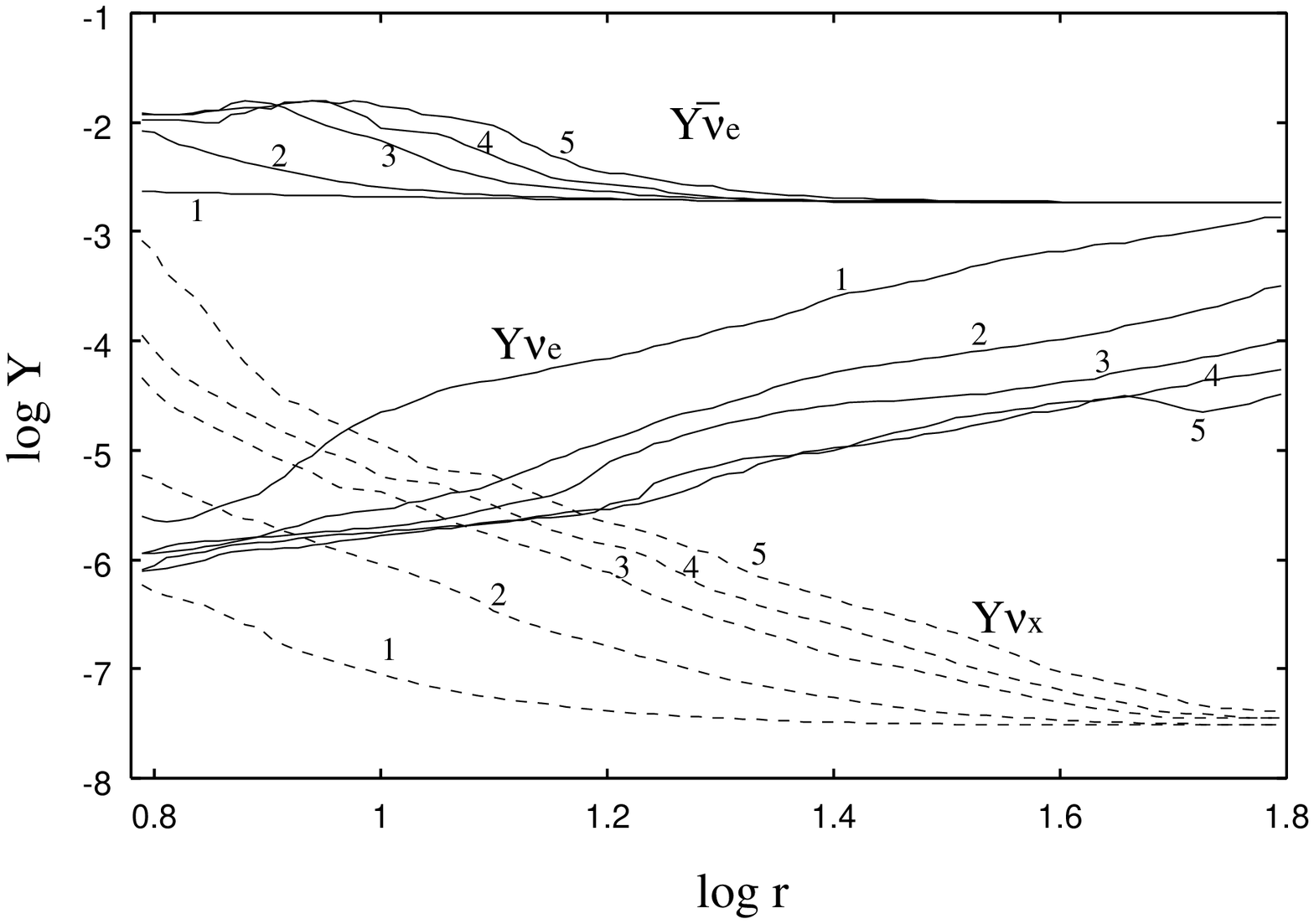}
\caption{The profiles of  of electron neutrino
 $Y_{\nu_e}$, electron anti-neutrino $Y_{\bar{\nu}_e}$, heavy neutrino
 $Y_{\nu_x}$.  Others are same as Fig.\ref{fig5}.} 
\label{fig6}
\end{figure}

\begin{figure}
\includegraphics[width=7cm,height=5cm]{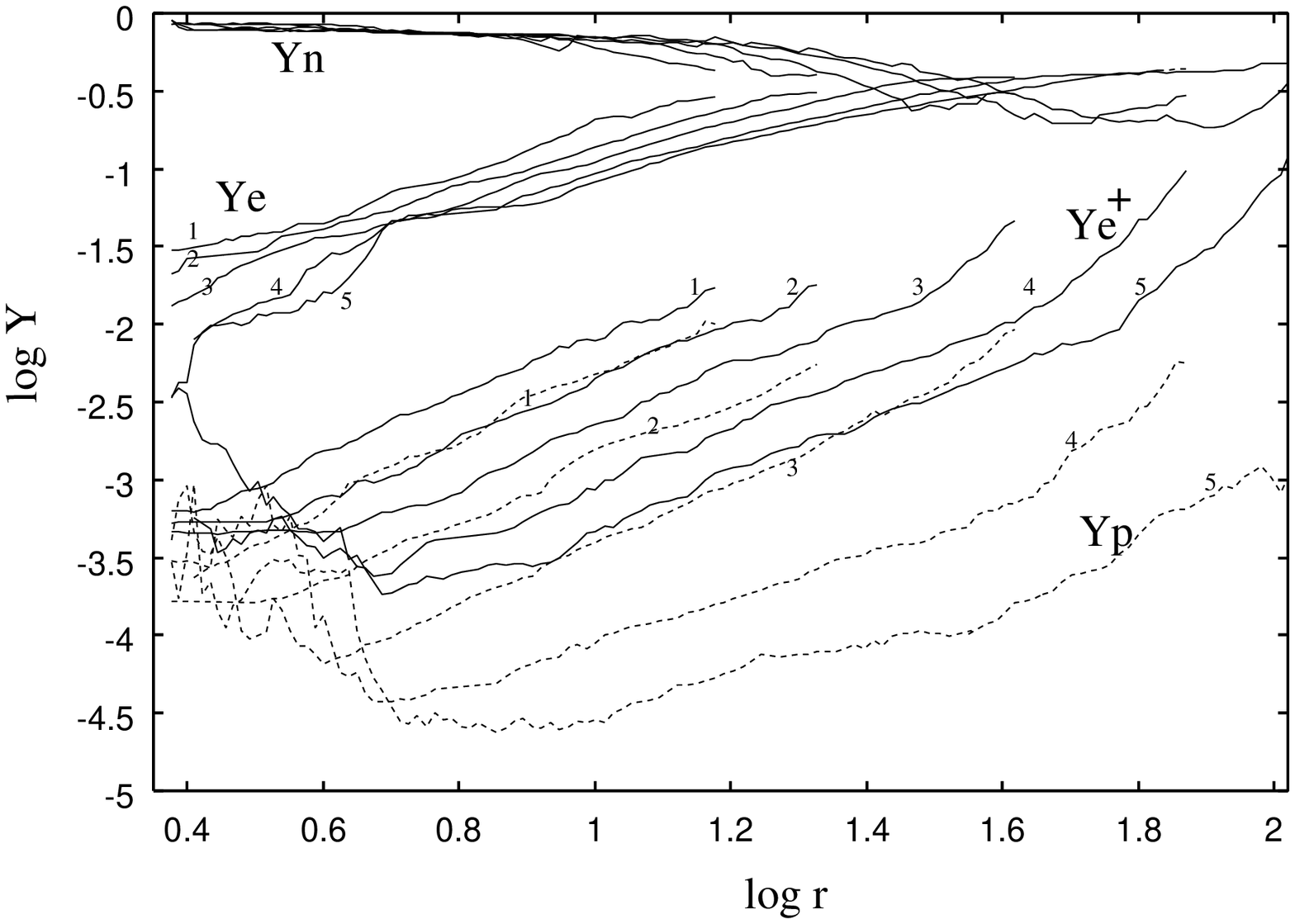}
\caption{The same as Fig.5, but for the cases of less accretion rates
with $a=0.9$ and $M_{BH}=3{\rm M}_\odot$. The numbers 
of the label 1,2,3,4,5 denote the 
 cases for accretion rates, $\dot{M}=$ 0.005,0.01,0.02,0.05,0.1(${\rm M}_\odot$ sec$^{-1}$).}
\label{fig7}
\end{figure}

\begin{figure}
\includegraphics[width=7cm,height=5cm]{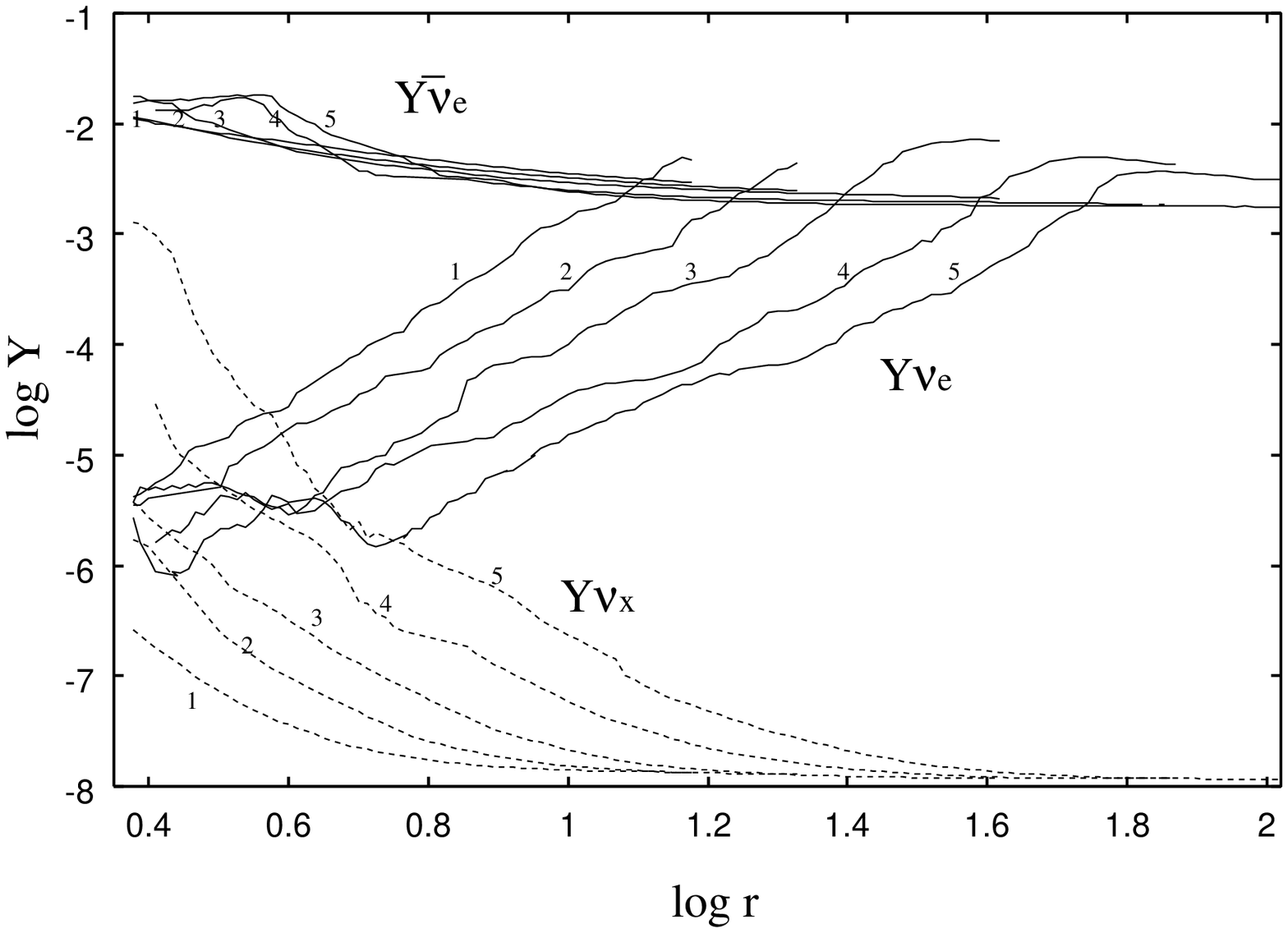}
\caption{The same as Fig.6, but for a rotating black hole with $a=0.9$. The numbers 
of the label 1,2,3,4,5 are same as in Fig. 7.}
\label{fig8}
\end{figure}

\subsection{Composition and chemical potential}
The composition of accreting matter which consists of free neutrino,
free proton, electron, positron, and each flavor type of neutrino is shown
in Fig. 5 -- 8. The massive accretion rates, $\dot{M}=0.05 \sim
1{\rm M}_\odot$sec$^{-1}$ for $a=0$, are selected in Fig. 5 and Fig. 6, while the less
ones, $\dot{M}=0.005 \sim 0.1{\rm M}_\odot$sec$^{-1}$ for $a=0.9$, are shown
in Fig. 7 and Fig. 8. The fraction of free neutron $Y_n$ increases along the accreting
flow and reaches to a dominant value at the inner side of the disk, $Y_n
\approx 0.7 \sim 0.8$.  However the fraction of free proton decreases
along the flow and reduces to an extremely small value, $Y_p \approx
10^{-3 \sim -4}$. When the accretion rate $\dot m$ increases, the proton
fraction decreases as, $Y_p \propto \dot{m}^{-1}$ though the fraction
of neutron remains a constant value at the inner side of the disk, $Y_n
\approx 0.7 \sim 0.8$. The number ratio of 
free neutron to free proton, $n_n/n_p$, is proportional to the
degeneracy factor, $\eta_{np}=(\mu_n-\mu_p)/kT$, i.e.,  $n_n/n_p \approx
\exp{(\eta_{np})}$. The locus of accreting matter in the $\rho
- T$ plane is traced to the increase of chemical potential of $\mu_n -
\mu_p$ (see Fig. 9). The larger accretion rate moves the locus into the
larger potential in $\mu_n - \mu_p$. In the typical case of the accretion its
factor $\eta_{np}$ increases from 5 at the outer boundary of the disk to
12 at the inner side of the disk (see Fig. 10). The accretion flow with large
efficiency of neutrino loss produces the locus with relatively low
temperature in the $\rho - T$ plane, $T \approx 2\sim 5 \times
10^{10}$ K, which means that the larger accretion flow has the larger number
ratio, $n_n/n_p$. These larger number ratios, $n_n/n_p \approx 10^{3\sim
4}$, result in the larger differences in the composition and opacity
between $\nu_e$ and $\bar{\nu}_e$.

The high dense matter and the sufficient time of reaction proceed the
electron capture on a proton. The electron fraction $Y_e$ gradually
decreases along the flow. The fraction of electron neutrinos,
$Y_{\nu_e}$, rapidly decreases along the accretion flow, while the
fraction of antielectron neutrinos, $Y_{\bar{\nu}_e}$, gradually
increases. These fraction changes are brought about mainly through the following reactions:
\begin{eqnarray}
e + p \rightleftharpoons \nu_e + n, \\
e^+ + n \rightleftharpoons \bar{\nu}_e + p, \\
e+e^+\rightleftharpoons \nu_e+\bar{\nu}_e
\end{eqnarray}
The fraction of heavy neutrinos $Y_{\nu_x}$ increases along the flow due
to the reactions, $e+e^+\rightleftharpoons \nu_i+\bar{\nu}_i$ and $N+N
\rightleftharpoons N+N+\nu_i+\bar{\nu}_i$. In some cases, $\dot{m} \ge
0.05$ for $a=0.9$, it exceeds that of $\nu_e$ at the inner side of disk,
$Y_{\nu_x} \ge Y_{\nu_e}$. The neutralization of the matter is
remarkable, $Y_e \le 0.03$ for $\dot{m} \ge 0.02$, near the
inside boundary, $r \approx r_{ms}$, where the equation of state for
dense, hot matter drastically changes. The composition of the accreting matter then
rapidly changes.

\begin{figure}
\includegraphics[width=9cm,height=6cm]{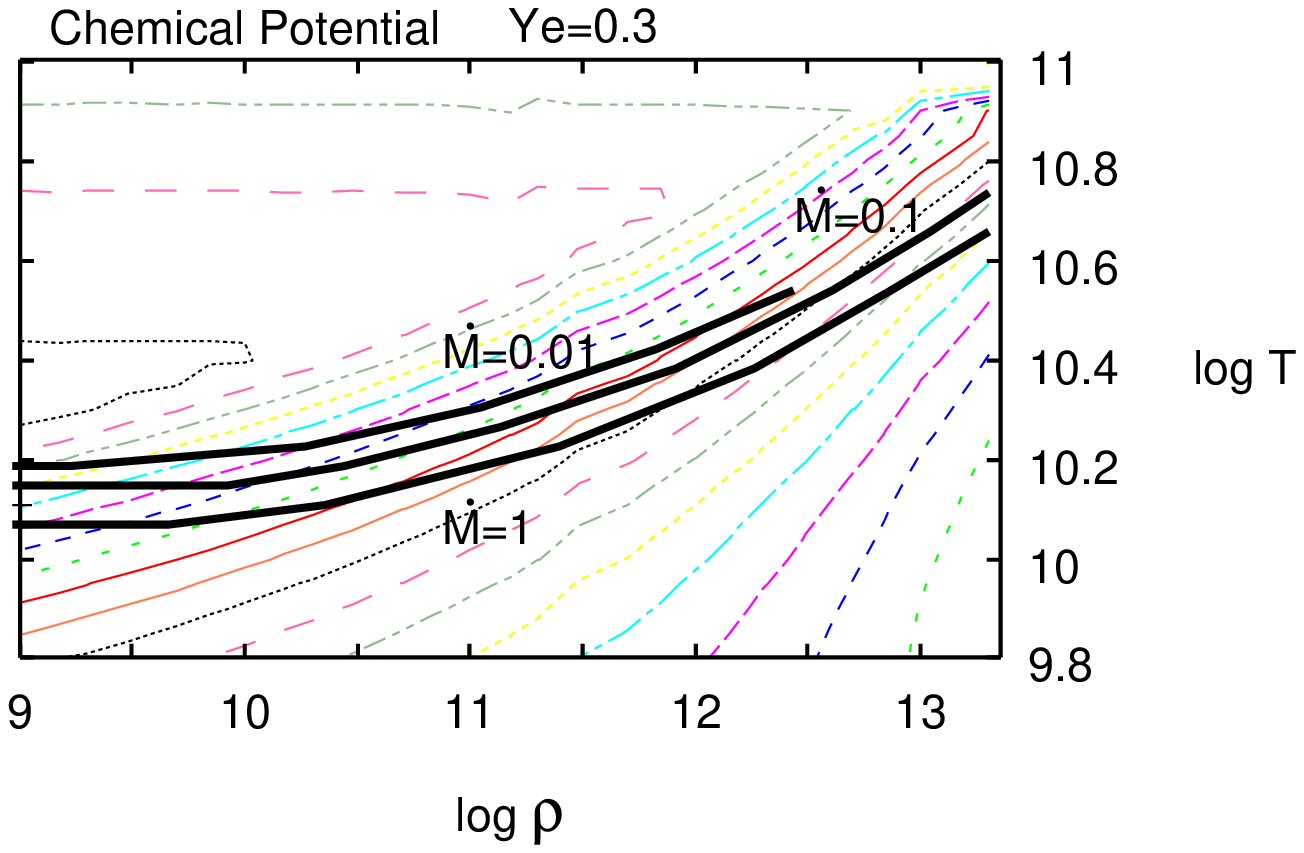}
\caption{The loci of accreting gas for the accretion rate
 $\dot{M}=0.01,0.1,1({\rm M}_\odot$ sec$^{-1}$) on the  $\rho$(g cm${^-3}$) --- $T$(K) plane. The contours
 of chemical potential $\mu_n - \mu_p$ with constant electron fraction 
 $Y_e=0.3$ are also plotted.}
\label{fig9}
\end{figure}

\begin{figure}
\includegraphics[width=7cm,height=5cm]{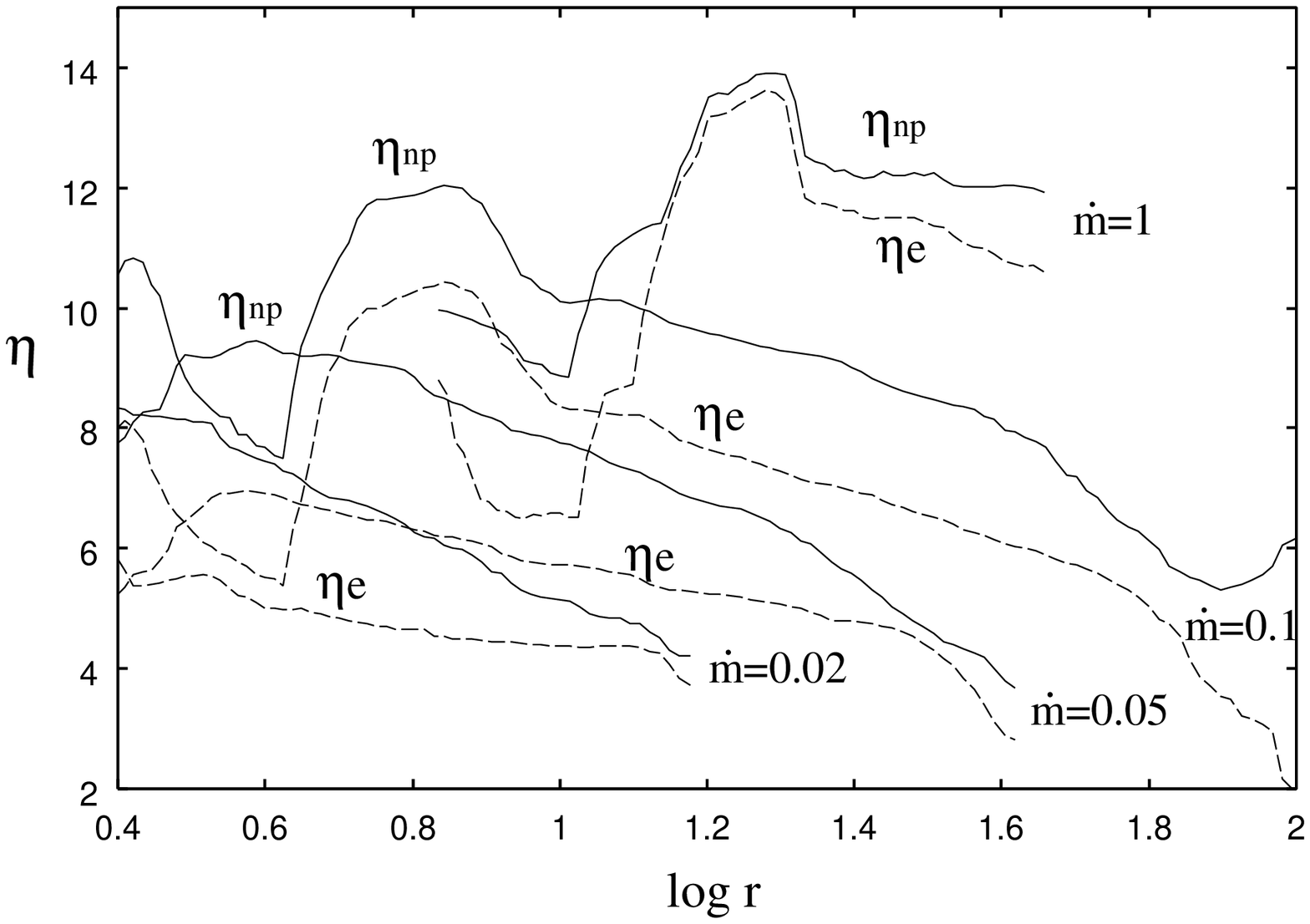}
\caption{The profiles of degeneracy factors for $\eta_{np}$ and
 $\eta_e$. The profiles denoted by the accretion rates $\dot
m$=0.1, 0.05, 0.02(M$_\odot$sec$^{-1}$) are shown, where $a=0.9$, and
those by $\dot m=1$(M$_\odot$sec$^{-1}$) are in the case with
$a=0$. The mass of a black hole is set to be $M_{BH}=3{\rm M}_\odot$.} 
\label{fig10}
\end{figure}

\subsection{Emissivities and opacities of neutrinos}
Let's evaluate the emissivities of neutrinos produced by the accretion.  
The emissivities by various types of reactions are shown in
Fig. 11, in the typical case of the accretion. The most efficient
emissivity is produced by the positron capture on 
 neutron, $e^+ + n \to \bar{\nu}_e + p$. At the outer side, $r \ge 20r_{ms}$, the pair
production, $e+e^+ \to 
\nu_e+\bar{\nu}_e$ and the nuclei reaction, $e + A \to \nu_e + A'$, also
contribute to the total emissivity.  At the inner side of the disk the URCA
process and nucleon---nucleon bremstrahlung become efficient. Though
the emissivity by antielectron neutrinos $q_{\bar{\nu}_e}$ is dominant,
the other type of neutrinos, ${\nu_e}, {\nu_x}$, 
 also produce some efficient emissivities. The emissivity by heavy
 neutrinos, $q_{\nu_x}$, becomes comparable with $q_{\bar{\nu}_e}$ near 
 the inner boundary, $q_{\bar{\nu}_e} \approx q_{\nu_x} \approx
 10^{34}$(erg/cm$^3$/sec). The emissivity increases sharply along the
flow and the total emissivity is expressed approximately as 
 $q = q_{\bar{\nu}_e}+ q_{\nu_e}+ q_{\nu_x} \approx 10^{36}
 r_*^{-4.25}$(erg/cm$^3$/sec).

\begin{figure}
\includegraphics[width=7cm,height=5cm]{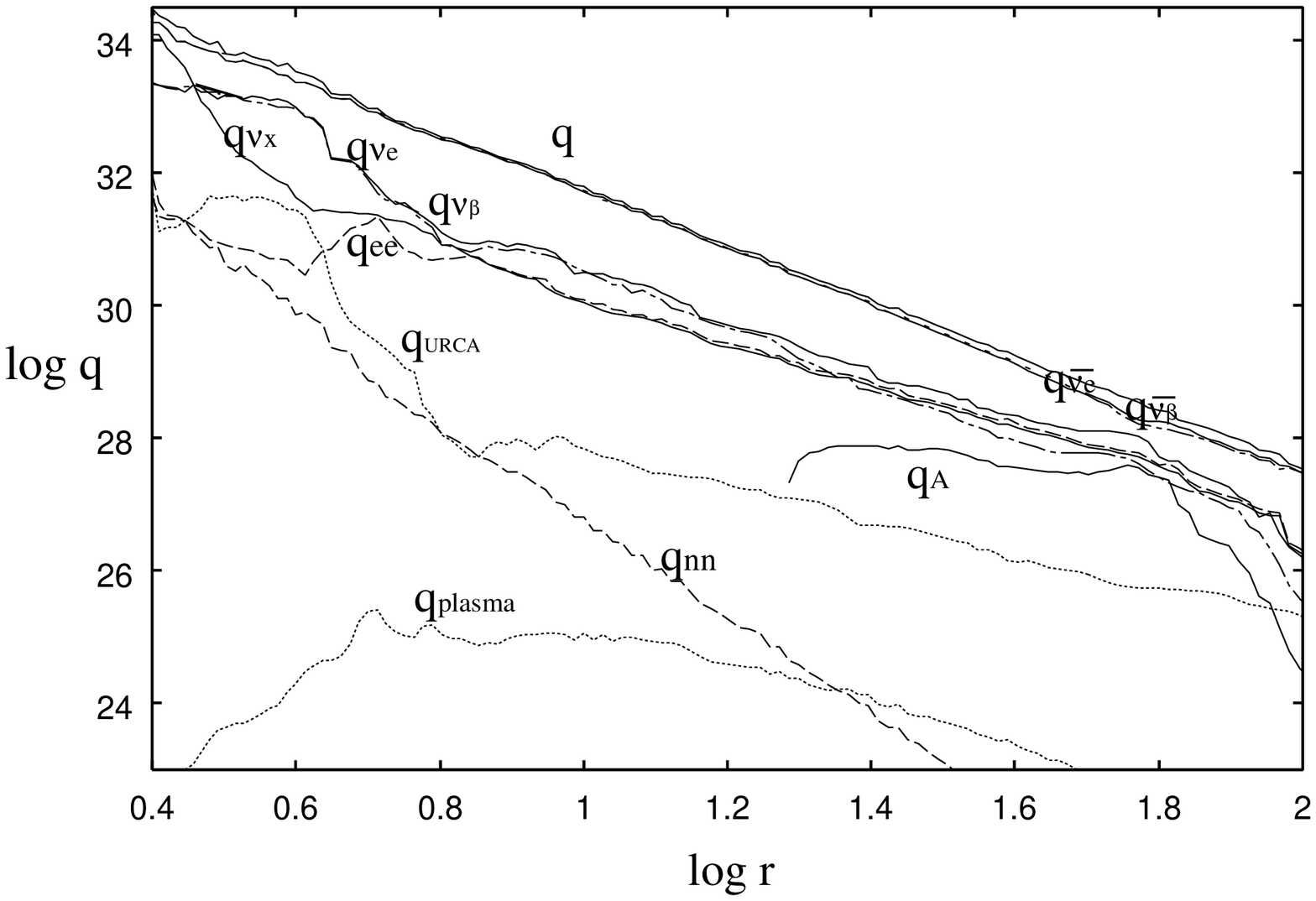}
\caption{The profile of total emissivity of neutrinos $q$ and those of antielectron 
neutrinos $q_{\bar{\nu}_e}$, electron neutrinos $q_{\nu_e}$, heavy neutrinos 
$q_{\nu_x}$ are plotted in the case of 
the accretion with $M_{BH}=3{\rm M}_\odot$, $a=0.9$, and
$\dot{M}=0.1{\rm M}_\odot$/sec. The emissivities by $\beta$ process  
$q_{\beta}$, positron capture on neutron $q_{e^+n}$, $e^-, e^+$ pair annihilation $q_{ee}$, URCA 
process $q_{URCA}$, nucleon-nucleon bremsstrahlung $q_{nn}$, plasma process 
$q_{plasma}$, and nucleus reaction $q_A$ are also depicted. The unit of
emissivity $q$ is (erg cm$^{-3}$sec$^{-1}$).}  
\label{fig11}
\end{figure}

\begin{figure}
\includegraphics[width=7cm,height=5cm]{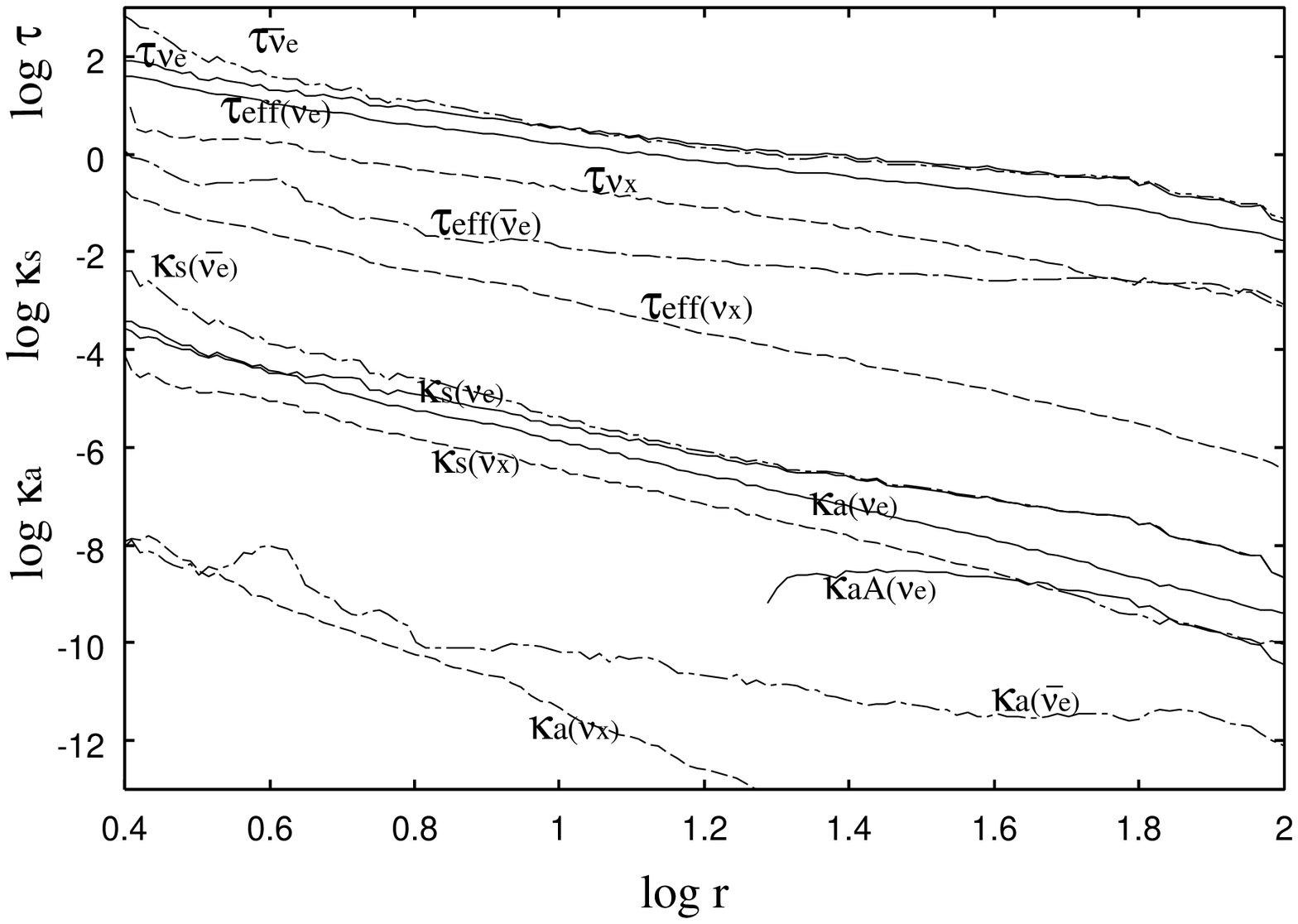}
\caption{The profiles of absorption opacity $\kappa_a(\nu_I)$, scattering opacity
 $\kappa_s(\nu_i)$, effective optical depth $\tau_{eff}(\nu_i)$, and total optical 
depth $\tau_{\nu_i}$, for $\nu_i=\nu_e, \nu_e, \nu_x$, are
plotted. The neutrino-nucleus opacity is labeled 
by $\kappa_{aA}$. The unit of
opacity $\kappa$ is (cm$^{-1}$). The parameters of accretion are same as in Fig. 11.}
\label{fig12}
\end{figure}

The opacities of absorption and scattering for each type of neutrinos are
shown in Fig. 12.  
The absorption opacity of $\nu_e$ is very much larger
than those of $\bar{\nu}_e$, $\nu_x$; 
$\quad \kappa_a(\nu_e) \gg \kappa_a(\bar\nu_e) \ge \kappa_a(\nu_x)$.  
The electron neutrino $\nu_e$ is absorbed mainly by dense free neutrons
through the reaction,
$\nu_e + n \to e + p $, while the antielectron neutrino $\bar{\nu}_e$
is little absorbed by rare free protons, $\bar{\nu}_e + p \to e^+ + n$.    
The scattering opacities of $\bar{\nu}_e$ and $\nu_e$ are comparable
except near the inside boundary where the chemical potential of
$\bar{\nu}_e$ is to some extent larger than that of $\nu_e$,
$\quad \eta_{\bar{\nu}_e} > \eta_{{\nu}_e}$, which brings about the
larger scattering opacity of  $\bar{\nu}_e$(see Table A2). The both total optical
depths of $\bar{\nu}_e$ and $\nu_e$, are very large with same order of
magnitude at the inside of the disk,
$\tau_{\bar{\nu}_e} =\tau_{\bar\nu_e a} +\tau_{\bar\nu_e s} \ge
\tau_{\nu_e}=\tau_{\nu_e a} +\tau_{\nu_e s} \sim 10^2$. However, the difference of effective optical depth 
between $\tau^{eff}_{\bar{\nu}_e}$ and $\tau^{eff}_{\nu_e}$ is remarkable, i.e.,
$\tau^{eff}_{\nu_e}(\ge 30) \gg \tau^{eff}_{\bar\nu_e}(\le 1)$, where the effective optical depth is defined by $\tau^{eff}_{\nu_i} = \sqrt{\tau^a_{\nu_i}
\tau^s_{\nu_i}}$. Most
of $\nu_e$ are absorbed in the disk and their energy is
deposited in the accreting matter. On the other hand the antielectron
neutrinos $\bar{\nu}_e$ carry out the thermal energy from the disk to
the outer atmosphere. The effective optical depth of $\nu_e$ is
approximately described as $\tau^{eff}_{\nu_e}\approx 6\times
10^3 r_*^{-2} (m)^{-1}\dot{m}$. It becomes over unity at the radius, $r
\le 14r_g(M_{BH}/3{\rm M}_\odot)^{-1/2}(\dot{M}/(0.1{\rm M}_\odot/ 
{\rm sec}))^{1/2}$, where the electron neutrino sphere(disk) is
formed. The sphere of antielectron neutrinos or that of heavy neutrinos are not formed when $\dot{m} \le 0.1$. The total optical depths of $\bar{\nu}_e$ and 
$\nu_e$ become over unity, $\tau_{\nu_e}, \tau_{\bar{\nu}_e} \ge 1$, where $r \le
 24r_g(M_{BH}/3{\rm M}_\odot)^{-1/2}(\dot{M}/(0.1{\rm M}_\odot/{\rm sec}))^{1/2}$.

\begin{figure}
\includegraphics[width=7cm,height=5cm]{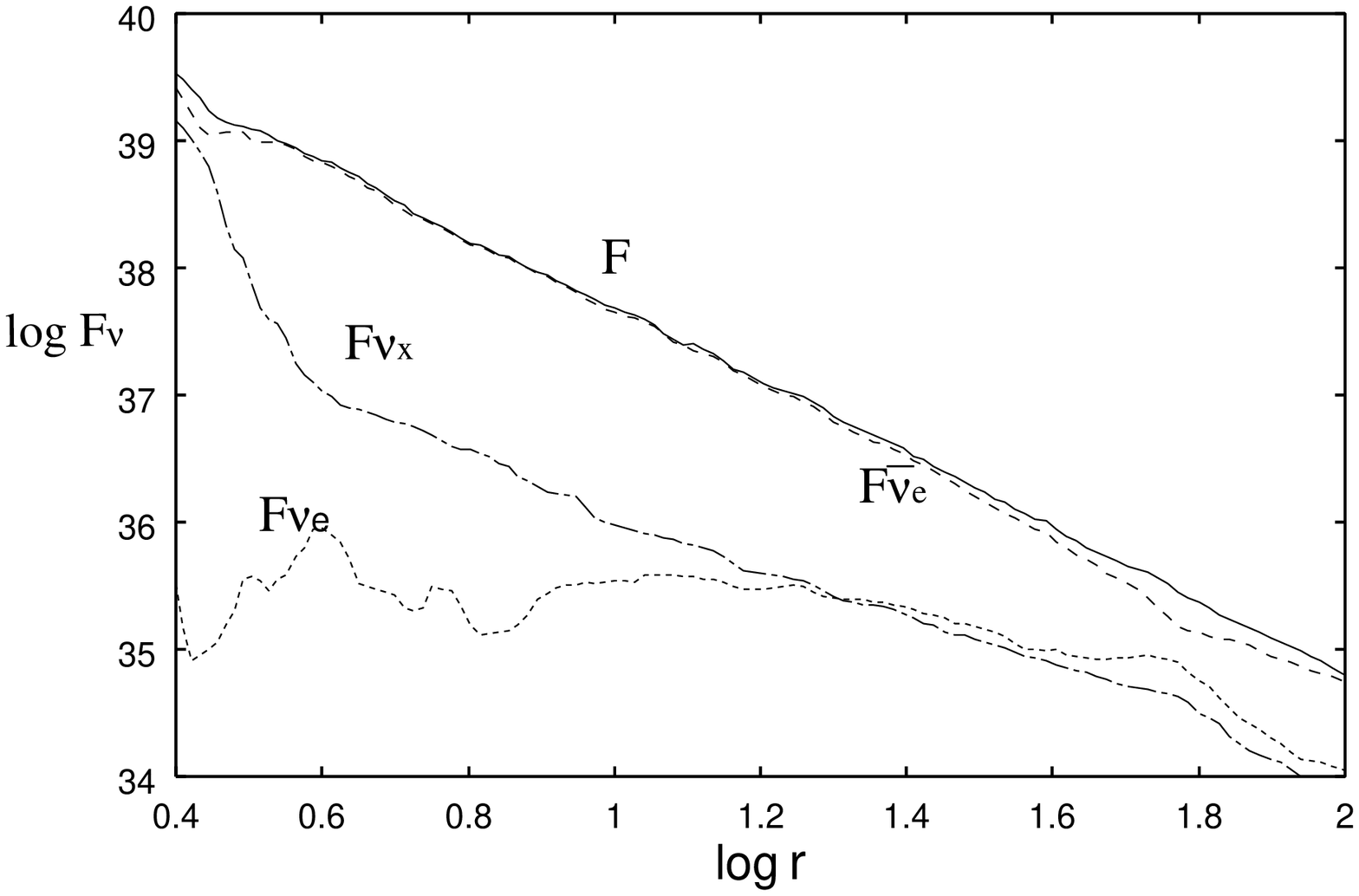}
\caption{The profiles of total flux density of neutrinos $F$, flux density of antielectron 
neutrinos $F_{\bar{\nu}_e}$, that of heavy neutrinos $F_{\nu_x}$, and that of 
electron neutrinos $F_{\nu_e}$. The unit of
flux density $F$ is  (erg cm$^{-2}$sec$^{-1}$). The parameters of
accretion are same as in Fig. 11.}
\label{fig13}
\end{figure}

The flux densities of neutrinos, $F_{\nu_i}(r)$, are 
shown in Fig. 11. The flux density of electron neutrinos, $F_{\nu_e}$, 
doesn't increase along the accreting flow. The ratio of the flux density
of $\bar{\nu}_e$ to that of 
$\nu_e$ becomes large to be $F_{\bar\nu_e} / F_{\nu_e} \approx 10^{3 \sim
4}$ at the inside of the disk.  The very large difference between
$F_{\bar\nu_e}$ and $F_{\nu_e}$ 
 seriously restricts the efficiency of the annihiration of $\nu_e$ and
 $\bar{\nu}_e$ at the outside of the disk, which results in the
 difficulty in the formation of a fire ball by the annihiration of $\nu_e, \bar\nu_e$ \citep{PWF}.    
 The flux density of $\nu_x$ rapidly increases 
near the inner boundary and reaches to be comparable with
$F_{\bar{\nu}_e}$. The profile of the total flux density is expressed approximately as
$F(r)=F_{\nu_e}+F_{\bar{\nu}_e}+F_{\nu_x} \approx 3.8\times 10^{40} r_*^{-2.9} ({\rm erg/cm}^2/{\rm sec})$.

\subsection{Luminosity and mean energy of neutrinos}
The angular momentum of a
black hole changes the efficiency of viscous heating through the shear stress $\sigma_{r\varphi}$. 
The energy equation is expressed such as
\begin{eqnarray}
&\frac{\partial E}{\partial t}& = Q^+ - Q^- \nonumber \\
 &=& 0.98\times 10^{43}({\rm 
erg/sec/cm}^2)\dot{m}m^{-2}r_*^{-3}  
(f_{vis}+f_{adv}+f_{\nu}) \nonumber \\
&=& 0,
\end{eqnarray}
where $f_{vis},f_{adv}$ and $f_{\nu}$ are 
the factors of energy flux by viscous heating, by advectional heating and by 
neutrino cooling. These factors, $f_{vis},f_{adv},f_{\nu }$, for $a=0$ and 
1 are shown in Fig. 14 and 15. The absolute values of $f_{vis}$ and
$f_{\nu}$ are orders of unity except for the inner side of the disk with
$a=1$. In the case of extreme Kerr, $a=1$, the efficiency of viscous heating $f_{vis}$ 
decreases to be $f_{vis}\approx 0.05 $ at the inner boundary of the
disk. The factor of advection, $f_{adv}$, is relatively small and positive,
$f_{adv}\approx 0.05$. The accretion with the
Keplerian angular momentum heats up the flow by the compression. On the
other hand the advection dominant accreting flow(ADAF) cools the flow by the extension of the fluid \citep{MPN,NPK,PWF}.

\begin{figure}
\begin{tabular}{cc}
\begin{minipage}{0.49\hsize}
\includegraphics[width=4cm,height=3cm]{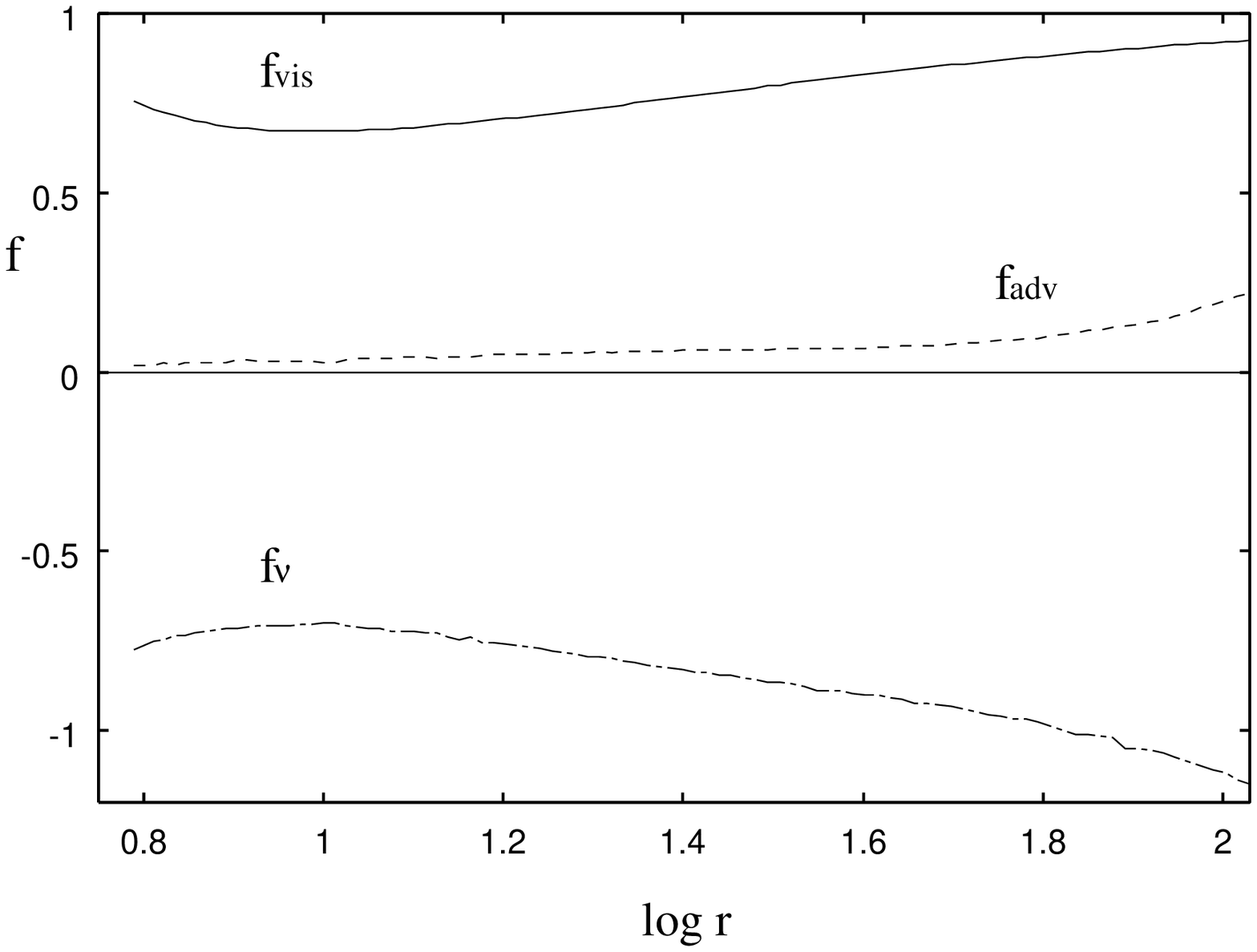}
\caption{The profiles of relative rates of heating and cooling. The relative heating 
rates due to viscosity $f_{vis}$, due to compression $f_{adv}$, and the relative cooling rate 
due to neutrino loss $f_\nu$ are plotted in the case with $a=0, M_{BH}=3{\rm M}_\odot, 
\dot{M}=0.1{\rm M}_\odot$ sec$^{-1}$.}
\label{fig14}
\end{minipage}&
\begin{minipage}{0.49\hsize}
\includegraphics[width=4cm,height=3cm]{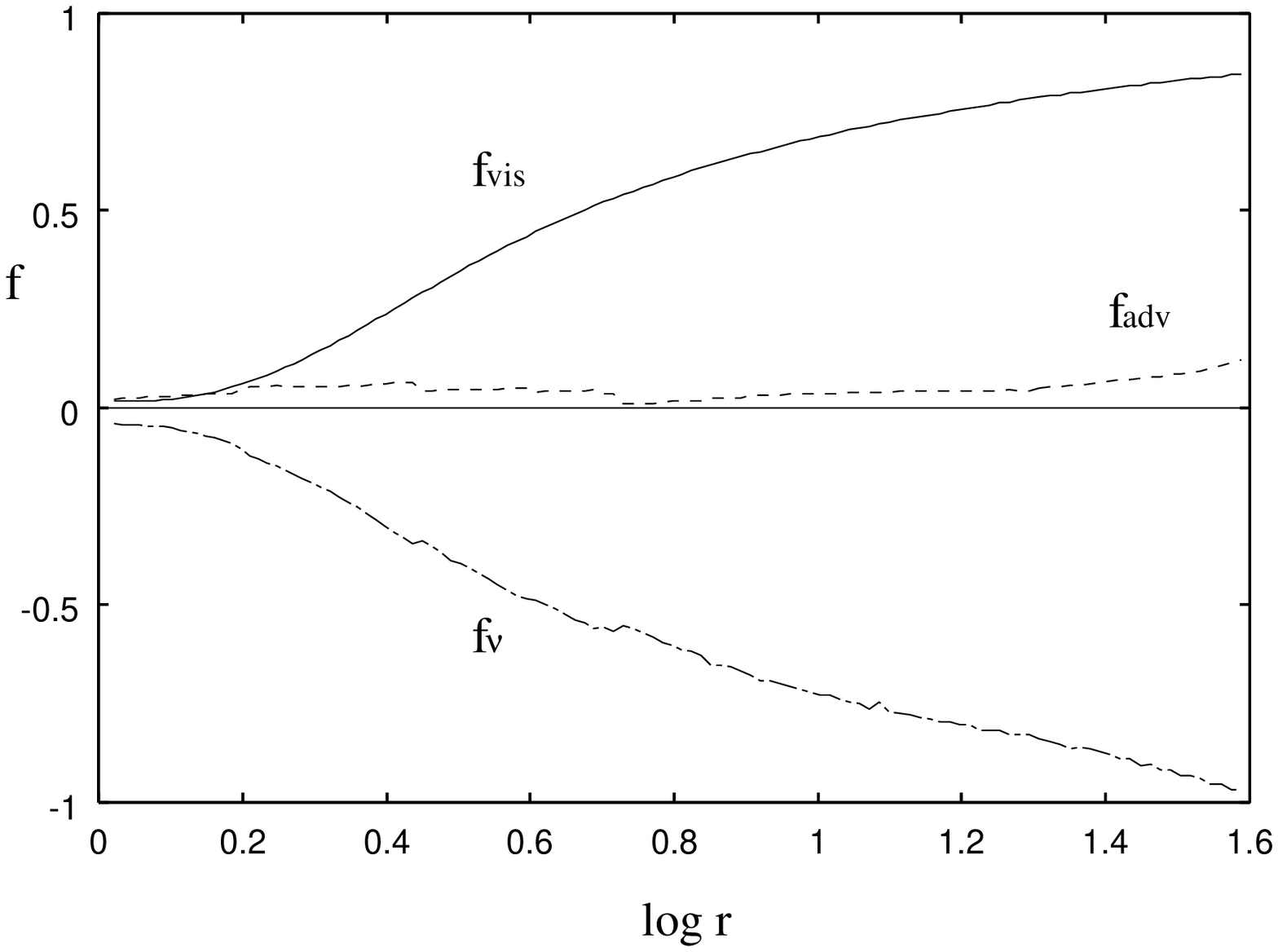}
\caption{The same as Fig.\ref{fig14}, but for the case of a rapidly
rotating black hole with $a=1$, $M_{BH}=3{\rm M}_\odot$  
 are depicted, where $\dot{M}=0.02{\rm M}_\odot$ sec$^{-1}$.}
\label{fig15}
\end{minipage}
\end{tabular}
\end{figure}

If the luminosity of each flavor of neutrino and its mean energy are
observed \citep{HH, McD,McK}, the physical structure around a black hole will be made clear. Though the electron neutrinos, $\nu_e$, are somewhat absorbed in the accreting matter at the outside of the disk, the antielectron neutrinos, $\bar{\nu}_e$, and heavy neutrinos, $\nu_x$, are little absorbed there. The energy flux of $\nu_e$ at the surface is very small in comparison with those of other types of neutrinos. Thus we calculate the luminosity $L_{\nu_i}$ and the mean energy
$\bar{E}_{\nu_i}$ for each type of neutrino $\nu_i$ 
emitted from the surface of the disk, and then discuss the
effects of the absorption of $\nu_e$ at the outer atmosphere of the disk if necessary. The luminosity measured at infinity is expressed in
the Boyer-Lindquist coordinate as $L_{\nu_i}=\int q_{\nu_i}^z(r) dS_z$, where
$q_{\nu_i}^z(r)$ is the $z$ component of the flux density vector at the radius
$r$, and $dS_z$ is the $z$ component of the surface element vector. The
component $q_{\nu_i}^z$ is connected with the local surface density $F_{\nu_i}$
such as $q_{\nu_i}^z=e^{\hat t}_t F_{\nu_i}$, where $e^{\hat t}_t$ is the $t$ component
of the orbiting basis
${\vec\omega}^{\hat t}$ (\ref{eq: one-form}). Thus the luminosity is
expressed by
\begin{equation}
L_{\nu_i}=2\pi \int_{r_h}^\infty \gamma (1+v_\omega v_{(\varphi)}) F_{\nu_i} rdr. 
\end{equation}

\begin{figure}
\includegraphics[width=8cm,height=6cm]{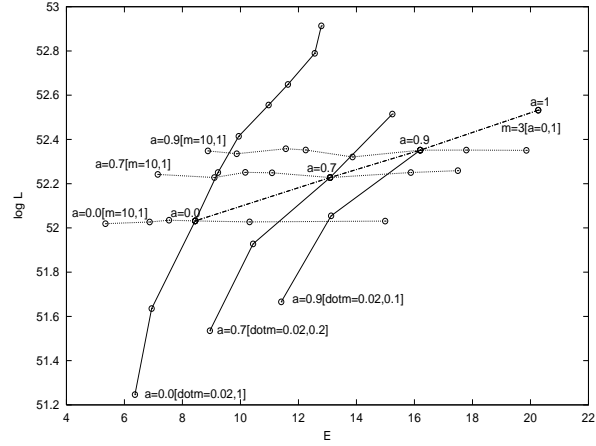}
\caption{The neutrino luminosity $L$(erg/sec) versus mean energy of ejected neutrinos 
${E}$(MeV) observed at infinity. The luminosity $L$ and mean energy
$E$ for the variation of specific angular momentum of a black hole, $a=0 \to 1$, are depicted by
the dot-dashed line labeled as $m=3[a=0,1]$, where 
the other parameters of accretion are fixed to be $M_{BH}=3{\rm M}_\odot, 
\dot{M}=0.1{\rm M}_\odot$/sec. Those for $M_{BH}=1,2,3,4,5,6,8,10 ({\rm M}_\odot)$
are connected by the line labeled as   
$a=0.9[m=10,1]$, where $a=0.9, \dot{M}=0.1{\rm M}_\odot$/sec. Those for
$\dot{M}=0.02,0.05,0.1 (\dot{M}_\odot$ sec$^{-1}$) are labeled as
$a=0.9[dotm=0.02, 0.1]$, where $a=0.9, M_{BH}=3{\rm M}_\odot$. Those 
for $a=0, 0.7$ are similarly shown.}  
\label{fig16}
\end{figure}

The mean energy $\bar{E}_{\nu_i}$ is defined as follows. When the
neutrinos are thermally equilibrium in the matter, 
the local mean energy of neutrinos $\bar{\epsilon}_{\nu_i}(r)$ 
is expressed by the matter temperature $T(r)$ and the degeneracy factor of
neutrino $\eta_{\nu_i}(r)$,  $\bar{\epsilon}_{\nu_i}(r) = kT(r) F_{3}(\eta_{\nu_i}(r))/F_{2}(\eta_{\nu_i}(r))$, where $F_k(\eta_{\nu_i})$ is the Fermi integral 
$F_k(\eta)\equiv \int_0^\infty{(1+\exp{(x-\eta)})^{-1}} {x^k dx}$. The
number per unit time emitted from the differential ring over the width of
radius $r \sim r+dr$ is
$dN_{\nu_i}(r) = dF_{\nu_i}(r)/\bar{\epsilon}_{\nu_i}(r)$, where
$dF_{\nu_i}(r)$ is the energy flux of neutrino from the ring. The total
number of neutrino is $N_{\nu_i}=\int_{r_h}^\infty dN_{\nu_i}(r)$. Thus the mean
energy is defined by $\bar{E}_{\nu_i}=L_{\nu_i}/N_{\nu_i}$. 

The total luminosity for all types of neutrinos is 
$L_\nu= \sum_i L_{\nu_i}$. The mean 
energy in all types of neutrinos is simply defined by $\bar{E}_\nu =
L_\nu/  \sum_i N_{\nu_i}$. 
We set the quantities of neutrinos within $r_{ms}$ to be same as the values at the 
boundary $r=r_{ms}$. The maximum radius of the integral is taken to be
$r_{max}=10^2r_{ms}$. 

The total luminosities and the mean energies measured at infinity, $L_\nu$, 
$\bar{E}_\nu$, are plotted in
Fig. 16 for some typical black holes with $m=1 \sim 10$ and $a=0
\sim 1$ and for the accretion rate $\dot m = 0.02 \sim 1$ . The rotating black hole has a small radius of the inner
boundary, $r_{ms}$, where the high density of accreting matter increases
the emissivity of neutrinos, which brings about the large luminosity
$L_\nu$ with high mean energy $\bar E_\nu$. It is superior to the
inefficiency of viscosity even if $a\to 1$ and $f_{vis} \to 0.05$. 
When the mass of a black hole is
$M_{BH}=3{\rm M}_\odot$ and the accretion rate is $\dot{M}=0.1{\rm M}_\odot/$sec,
the ratio of luminosity is
$L_\nu(a=1)/L_\nu(a=0) \approx 3.5$, and that of mean energy is
$\bar E_\nu(a=1)/\bar E_\nu(a=0) \approx 2.5$. The luminosity is proportional to
the accretion rate, $L_\nu \propto \dot{M}$, but is  
independent of the mass scale of a black hole, $M_{BH}$, that is, the constant ratio 
of the gravitational binding energy of accreting matter is released into the 
neutrino's energy through the viscous heating. Its ratio increases according to $a$. 
The mean energy $\bar E_\nu$ is proportional to $\dot{M}$ but is inversely proportional 
to $M_{BH}$, $\bar E_\nu \propto \dot{M}^{0.8} M_{BH}^{-1.07}$. 
When each type of neutrino ejected from a GRB is observed, the physics
around a black  hole will be made more clear. The observed
luminosities of neutrinos and their mean energies will determine the
physical parameters of a central black hole, $a, M_{BH}$, and the
accretion rate, $\dot M$, from the diagram shown in Fig. 16. 

\section{Discussion}
We have studied the accretion in simplified model in which the disk is
assumed to be stationary and its vertical structure is treated by
one-zone model with the thickness, $H$. These simplifications might
restrict the properties 
of the accretion disk. We first discuss the thermal stability of the
disk. In the stationary state, the cooling rate due to neutrino's loss
must be equal to the production rate of thermal energy in the disk
even if neutrinos are scattered in many times within the disk. 
The accretion disk is thermally stable since the antielectron
neutrinos $\bar{\nu}_e$ are little absorbed in the disk and can
sufficiently  
carry out the perturbed thermal energy. 
However in the short duration within the diffusion time, $t_{diff}
\approx \tau_s H/c \approx 10$(ms), the perturbed energy is stayed
in the disk. The heating or cooling time then becomes 
$t_{heat}= t_{cool}= \approx \epsilon/q \approx
$a few ms, i.e., $t_{heat} <t_{diff}$. If the accretion disk is
thermally unstable, the perturbations rapidly grow. This
stability should be decided by the fully nonstationary treatment of 
accretion disk. Here we discuss it in brief by the perturbation
analysis.
When the scattering optical depth is large $\tau_s \gg 1$ and the
absorbing optical depth is negligible, $\tau_a \ll 1$, the neutrinos
stay  
in the disk in $\tau_s$ times of the crossing time
$t_{cross}=H/c$. When $\tau_s \ll 1$, the perturbation of the cooling
rate is expressed as $\delta {Q}^- = \delta F_{\nu} = \delta (q_\nu
H)$. We provide the perturbation of the cooling rate in the thick case
as $\delta {Q}^- = \delta (q_\nu H/\tau_s)$. The perturbed cooling
rates by $\bar{\nu}_e$ and ${\nu}_x$ are then set to be
$\delta{Q}_{\bar{\nu}_e}^- + \delta{Q}_{\nu_x}^- = \delta
(q_{\bar{\nu}_e}H /(1+\tau_s(\bar{\nu}_e)))+ \delta (q_{\nu_x}H
/(1+\tau_s(\nu_x)))$. Since the electric neutrinos $\nu_e$ are almost 
absorbed, the perturbed cooling rate is expressed as
$\delta{Q}_{\nu_e}^-=\delta({q_{\nu_e}H}/(1.5\tau_{\nu_e
a}\tau_{\nu_e} +\sqrt{3}\tau_{\nu_e a}+1))$.

\begin{figure}
\includegraphics[width=7cm,height=5cm]{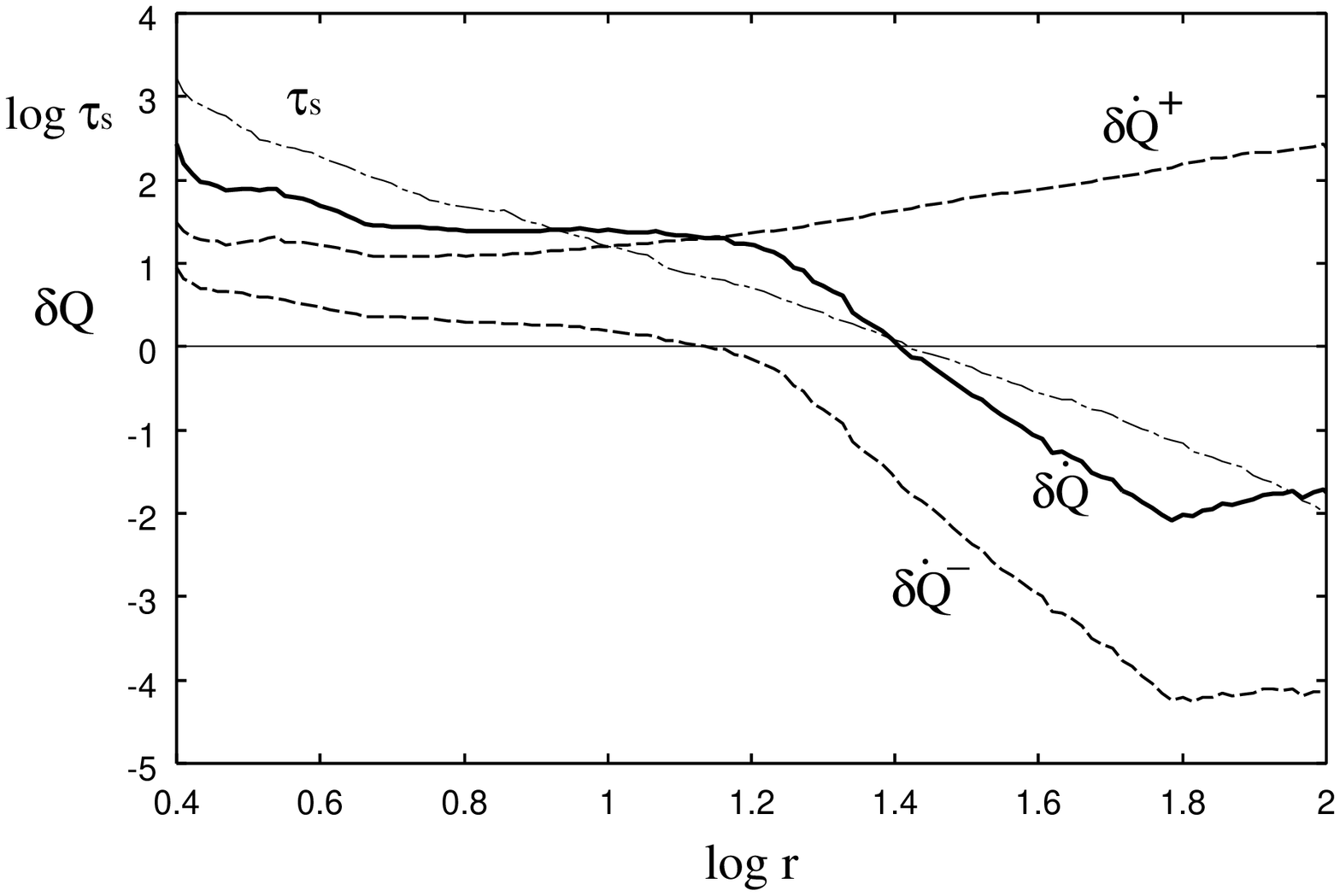}
\caption{The variation of net heating rate $\delta \dot{Q}$ due to the
 increase of temperature $\delta T$ within the diffusion time. The
 variation of the rate by the viscous and compressing heating
 $\delta \dot{Q}^{+}$, and that by the neutrino cooling $\delta 
\dot{Q}^-$ are plotted by dashed lines. The scattering optical depth of antielectron 
neutrinos $\tau_s$ is plotted by dot-dashed line. The parameters of accretion 
are $\dot{M}=0.5{\rm M}_\odot{\rm sec}^{-1}, M_{BH}=3{\rm M}_\odot, a=0.9, \alpha_{vis}=0.1$.} 
\label{fig17}
\end{figure}

The perturbation of the heating rate is brought about by the viscous
heating and the compression, $\delta {Q}^+ = \delta
{Q}_{vis}+\delta{Q}_{adv}$. We thus evaluate the thermal stability for
the perturbation of temperature $\delta T$ with the restriction of
constant surface density $\Sigma$,
\begin{equation}
\delta{Q}=\biggl(\frac{\partial\log{{Q}^+}}{\partial\log{T}}\biggr)_\Sigma
+ \biggl(\frac{\partial \log{Q}^-}{\partial\log{T}}\biggr)_\Sigma. 
\end{equation}
 
The results for $\delta{Q}$ are shown in Fig. 17, where $\dot{m}=0.5,
m=3, a=0.9 and \alpha_{vis}=0.1$. The sign of $\delta{Q}$ becomes
positive when $\tau_s(\bar{\nu}_e) \ge 1$. The sign of $\delta
{Q}^-=({\partial\log{ Q^-}}/{\partial\log{T}})_\Sigma$ also becomes positive
when $\tau_s(\bar{\nu}_e) \ge 10$, since the opacity of $\bar{\nu}_e$
increases according to the rise in temperature, $\delta T>0$. For the region where 
the scattering optical depth of $\bar{\nu}_e$ is larger than unity,
$\tau_s(\bar{\nu}_e) \ge 1$, the  
accretion disk becomes thermally unstable at least within the duration of the diffusion time. 

We next discuss the interaction between neutrinos and nucleons. The spectrums of 
neutrino are assumed here to be Ferm-Dirac distributions with local temperature being 
equal to that of nucleon medium. In the neutrino sphere(disk) the neutrinos are thermally 
equilibrium with nucleons but in the outer layer above it the above assumption is 
in general not consistent. In the outer layer the spectral temperature defined by 
the mean energy of neutrinos, $T_* \equiv \bar{E}_\nu /3$, becomes higher than the 
nucleon's temperature \citep{R01}. Since the mean energy of neutrino exceeds the rest 
mass energy of electron, $\bar{E}_\nu > m_ec^2$, the collision between neutrinos and 
electrons effectively transfer the momentum and energy from neutrinos
to electrons \citep{TBH}. The dynamic structure functions which include the full 
kinematics of $\nu_i$-nucleon scattering produce also the tranfers of energy and 
momentum between neutrinos $\nu_i$ and nucleons. In the diffusion zone in which the 
neutrinos many scatter the electrons and nucleons, $\tau_s \ge 1$, the convection 
originated by neutrinos may work important effects on the dynamics of the medium. 
In our accretion model the effective optical depth, $\tau_{eff} \equiv \sqrt{\tau_a 
\tau_s}$, exceeds unity only for $\nu_e$. The thermal equilibrium disk of $\nu_e$ is 
formed at $r \approx 10r_g$. The scattering optical depths of all flavors, 
$\tau_s({\nu_e}), \tau_s({\bar{\nu}_e}), \tau_s({\nu_x})$, exceed
unity. In the diffusion zone above the disk, $z \ge H$, the neutrinos
might push the nucleon matter outward and might drive the convection or wind
or jets. 

Our results for the hot, dense matter accretion onto a black hole indicate
that the emissivity of electron antineutrinos $q_{\bar{\nu}_e}$ is very
efficient and the fraction of free protons is extremely minute, which allows
the accretion disk effectively cooled by $\bar{\nu}_e$. On the other
hand the electron neutrinos $\nu_e$ are almost trapped by many free
neutrons. The accretion disk cooled by neutrinos could be thermally
unstable at least within the diffusion time,
$t_{diff}\approx 10$(ms). Our treatment of the accretion is very
simple, stationary model with the homogeneous disk in the vertical
direction. The compactness problem for GRBs requires the formation of
relativistic jets. The accretion should be further solved in
unstationary, 2- or 3- dimensional treatment with the 
transfers of neutrinos.


\section*{Acknowledgments}

I thank Professor T. Yoshida for helpful discussions,
 Professor S. Yanagita for some helpful suggestions and an anonymous
 referee for very useful comments that improved 
the presentation of the paper.





\appendix
\section{Reaction rates of neutrinos}
The neutrino cross
sections, opacities, emissivities and reaction rates used in our model
are listed in Table A1--4. The numerical functions used in the reaction rates are listed in Table A5.

\begin{table*}{p}
\begin{minipage}{160mm}
\caption{Cross section}
\label{tbl-2}
\begin{tabular}{@{}lll}
No. & Neutrino Reaction & Cross Section\\
\hline
1 &$\nu_e + n \to e + p$&$\sigma_{\nu_e, n}^a(\varepsilon_{\nu_e}) = 
\sigma_0(1+3g_A^2)\Bigl(\frac{\varepsilon_{\nu_e}+Q}{2m_e c^2}\Bigr)^2[1- 
\Bigl(\frac{m_e c^2}{\varepsilon_{\nu_e}+Q}\Bigr)^2]^{1/2}W_M$\\
2 &$\bar{\nu}_e + p \to e^+ + n$&$\sigma_{\bar{\nu}_e, 
p}^a(\varepsilon_{\bar{\nu}_e}) = 
\sigma_0(1+3g_A^2) \Bigl( \frac{\varepsilon_{\bar{\nu}_e}-Q}{2m_e c^2}
\Bigr)^2 [1-\Bigl(\frac{m_e c^2} {\varepsilon_{\bar{\nu}_e}-Q}\Bigr)^2
]^{1/2} W_{\bar{M}}$\\
3 &$\nu_i+p \to \nu_i+p $&$\sigma_{p}^s(\varepsilon_{\nu_i}) = \sigma_0 
\Bigl(\frac{\varepsilon_{\nu_i}}{2m_e 
c^2}\Bigr)^2[4\sin^4\theta_W-2\sin^2\theta_W+ \frac{1+3g_A^2)}{4}]$\\
4 &$\nu_i+n \to \nu_i+n $&$\sigma_{n}^s(\varepsilon_{\nu_i}) = \sigma_0 
\Bigl(\frac{\varepsilon_{\nu_i}}{2m_e c^2}\Bigr)^2\Bigl(\frac{1+3g_A^2)}{4}\Bigr)$\\ 
5 &$\nu_i+e \to \nu_i+e $&$\sigma_e = \sigma_0\frac{1}{8}\Bigl[(C_V+C_A)^2 + 
\frac{1}{3}(C_V-C_A)^2\Bigr] 
\Bigl(\frac{\varepsilon_{\nu_i}}{m_ec^2}+\frac{1}{2} \Bigr)$  \\
\hline
\end{tabular}
\footnote{These cross sections are given by  Burrows(2001). A
 convenient reference neutrino cross section is 
$\sigma_0 = \frac{4G_F^2(m_ec^2)^2}{\pi(h c/2\pi)^4}\simeq 1.705\times 
10^{-44}{\rm cm}^2$. $g_A$ is the axial-vector coupling constant ($\sim - 
1.26),$  $\theta_W$ is the Weinberg angle and $\sin^2\theta_W \simeq 0.23$, 
$Q=m_nc^2-m_pc^2=1.29332$MeV, and for a collision in which the electron gets all of 
the kinetic energy $\varepsilon_{e^-}=\varepsilon_{\nu_e}+Q$. $W_M$ is the weak 
magnetism/recoil correction and is approximately equal to 
($1+1.1\varepsilon_{\nu_e}/m_nc^2$). $C_V=1/2+2\sin^2\theta_W$ for $\nu_e$ and 
$\bar{\nu}_e$ neutrinos, $C_V=-1/2+2\sin^2\theta_W$ for $\nu_x$ and 
$\bar{\nu}_x$ neutrinos, $C_A=+1/2$ for $\nu_e, \bar{\nu}_\mu$ and 
$\bar{\nu}_\tau$ neutrinos, and $C_A=-1/2$ for $\bar{\nu}_e, {\nu}_\mu$ and 
${\nu}_\tau$ neutrinos.}
\end{minipage}
\end{table*}

\begin{table*}{t}
\begin{minipage}{160mm}
\caption{Opacity}
\label{tbl-3}
\begin{tabular}{@{}llll}
No. & {Neutrino Reaction} & {Opacity}  & {References} \\
\hline
1 &$\nu_e + n \to e + p$&$\kappa^a_\beta = \eta_{np}\frac{\int d^3p 
f_{\nu_e}(\varepsilon) \sigma_{\nu_e, 
n}^a(\varepsilon)[1-f_{e}(\varepsilon_{\nu_e}+Q)]}{\int d^3p 
f_{\nu_e}(\varepsilon)}$& Burrows(2001)  \\
  &$ $&$
\approx 4.3\times 
10^{-6}({\rm cm}^{-1})\rho_{12}Y_{np}T_{11}^2\frac{F_{4}^+(\eta_{\nu_e},\eta_e,q)}{F_2(\eta_{\nu_e})}
 $&   \\
2 &$\bar{\nu}_e + p \to e^+ + n$&$\kappa^a_{\bar\beta}\approx 4.3\times 
10^{-6}({\rm cm}^{-1})\rho_{12}Y_{pn}T_{11}^2\frac{F_{4}^+(\eta_{\bar{\nu}_e},\eta_{e^+},-q)}{F_2(\eta_{\bar{\nu}_e})} $ & Burrows(2001)  \\
3 &$\nu_i+\bar{\nu}_i + N+N \to N+N $&$\kappa^a_{NN}\approx 
\frac{3C_A^2G_F^2n_BT^5\Gamma}{\pi^2}\frac{2}{5\varepsilon_\nu}$ &Raffelt(2001) 
\\ 
  &$  $&$\approx 1.43\times 
10^{-9}({\rm cm}^{-1})(X_n\rho_{12})^2T^2_{11}\frac{F_2(\eta_{\nu_i})}{F_3(\eta_{\nu_i})}\Bigl(
\frac{3}{2+0.862T_{11}}\Bigr)^{1/2}
  $ &  \\
4 &$\nu_e+A \to e+A'$&$\kappa^a_A = \frac{\alpha^2}{4}\sigma_0 n_A e^{\eta_n -\eta_p 
-q'} \frac{2}{7} N_p(Z)N_h(N)$ &  \\
 & &$  \Bigl(\frac{kT}{m_e c^2} \Bigr)^2 
\frac{F_4^{q'-}(\eta_{\nu_e},\eta_e)}{F_2(\eta_{\nu_e})} $  & Bruenn(1985)  \\
  &$    $& $\approx 1.2\times 10^{-6}({\rm cm}^{-1})\rho_{12} Y_A T_{11}^2 f_A(\eta)$
 &   \\
5 &$\nu_i+N \to \nu_i+N $&$\kappa^s_N \approx 1.1\times 
10^{-6}({\rm cm}^{-1})\rho_{12}T_{11}^2C_{s,N}Y_{NN} 
\frac{F_4(\eta_{\nu_i})}{F_2(\eta_{\nu_i})} 
 $ & Burrows(2001) \\
6 &$\nu_i+e \to \nu_i+e $&$\kappa^s_e \approx 1.66\times 10^{-6}({\rm cm}^{-1}) Y_e\rho_{12} 
T_{11}^2  \frac{F_4(\eta_{\nu_i})}{F_2(\eta_{\nu_i})}   $ & Burrows(2001)  \\
7 &$\nu_i + A \to \nu_i+ A$&$\kappa^s_A=\frac{a_0^2}{6} \sigma_0 n_A A^2 
\Bigl( \frac{kT}{m_ec^2} \Bigr)^2 \frac{F_4(\eta_{\nu_i})}{F_2(\eta_{\nu_i})} $& 
Bruenn(1985) \\ 
  &  &$\approx 2.3\times 10^{-7}({\rm cm}^{-1}) \rho_{12} T_{11}^2 A^2 Y_A  
\frac{F_4(\eta_{\nu_i})}{F_2(\eta_{\nu_i})}$ &    \\
\hline
\end{tabular}
\footnote{The absorption opacity $\kappa_a$ and scattering opacity
 $\kappa_s$ are expressed with unit cm$^{-1}$. $F_k(\eta)$ is the Fermi integral 
$F_k(\eta)\equiv \int_0^\infty \frac{x^k dx}{1+\exp{(x-\eta)}}$. The complex 
integrals, $F_{4}^+(\eta_{\nu_e},\eta_e,q)$ and $F_4^{q'-}(\eta_{\nu_e},\eta_e)$, 
are shown in Table A5. The doubly indexed quantity $Y_{np}$ is the number fraction 
defined as $Y_{np}=\eta_{np}/n_B$, and $Y_{pn}$ is $Y_{pn}=\exp{(\eta_p - 
\eta_n)}Y_{np}$, where $\eta_{np}$ is the dynamic structure factor for 
neutrino-neucleon interaction, $
\eta_{np} = \int 
\frac{2d^3p}{(2\pi)^3}f_n(\varepsilon)[1-f_p(\varepsilon)]=\frac{n_p-n_n}{\exp{(
\eta_p -\eta_n)} -1} 
$. $Y_{NN}$ is $Y_{NN}=\frac{Y_N}{1+\frac{2}{3}\max{(\eta_N, 0)}}$,
and $C_{s,N}$ is $C_{s,N}=1$ if $N=n$ and $C_{s,N}=0.827$ if $N=p$. $A$
 is the mean mass number of heavy nuclei, $n_A$ is the number density
 of heavy nuclei and $Y_A$ is the fraction, $Y_A=n_A/n_B$. $X_n$ is the mass fraction 
of neutrons. 
$f_A(\eta)$ is $f_A(\eta)=e^{\eta_n -\eta_p -q'} \frac{2}{7} N_p(Z)N_h(N) 
\frac{F_4^{q'+}(\eta_{\nu_e},\eta_{e})}{F_2(\eta_{\nu_e})}$, where 
$q'\approx \eta_n-\eta_p+\Delta, \quad \Delta=\frac{3{\rm MeV}}{kT}$. 
$N_p(Z)$ and $N_h(N)$ are 
$N_p(Z)=\{0 \ {\rm for} \ Z<20;\ Z-20 \ {\rm for} \ 20<Z<28; \ 8 \ {\rm for}\ Z>28\}$, 
and 
$N_h(N)=\{6 \ {\rm for} \ N<34;\ 40-N \ {\rm for} \ 34<N<40; \ 0 \ {\rm for} \ N>40\}$. 
The density $\rho_{12}$ and temperature $T_{11}$ are normalized vales
by $10^{12}$(g/cm$^3$) and $10^{11}$K}.
\end{minipage}
\end{table*}

\begin{table*}
\begin{minipage}{160mm}
\caption{Emissivity}
\label{tbl-4}
\begin{tabular}{@{}llll}
No. & {Neutrino Reaction} & {Emission Rate}  & {References} \\
\hline
1 &$ e + p \to \nu_e + n$&$q_\beta(\nu_e)= \eta_{pn} \frac{4\pi c}
{(h c)^3}\int_0^\infty d^3p \varepsilon f_{e}(\varepsilon+Q) \sigma_{\nu_e, 
n}^a(\varepsilon)[1-f_{\nu_e}(\varepsilon)]$& Burrows(2001)  \\
  &$ $&$\approx 7.4\times 
10^{33}({\rm erg cm}^{-3}{\rm sec}^{-1})\rho_{12}T_{11}^6Y_{pn}F_{5}^-(\eta_{\nu_e},\eta_e,q)  $&   \\
2 &$e^+ + n \to \bar{\nu}_e + p$&$q_{\bar\beta}(\bar{\nu}_e) \approx 7.4\times 
10^{33}({\rm erg cm}^{-3}{\rm sec}^{-1})\rho_{12}T_{11}^6Y_{np}F_{5^+}^-(\eta_{\bar{\nu}_e},\eta_{e^+},q)
$ &  
Burrows(2001)  \\
3 &$ N+N  \to N+N+\nu_i+\bar{\nu}_i$&$q_{NN}(\nu_i,\bar{\nu}_i) \approx 7.3\times 
10^{30}({\rm erg cm}^{-3}{\rm sec}^{-1}) (X_N\rho_{12})^2 T_{11}^{5.5}
  $ &  Burrows(2001)  \\
4 &$e+A \to \nu_e + A'$&$q_A(\nu_e) = b\sigma_0 \frac{\alpha^2}{4} N_A \frac{2}{7} 
N_p(Z) N_h(N)\Bigl(\frac{kT}{m_e c^2} \Bigr)^5 kT 
F_5^{q'-}(\eta_{\nu_e},\eta_{e})$ & Bruenn(1985)  \\
 &  &$\approx 2.06\times 10^{33}({\rm erg cm}^{-3}{\rm sec}^{-1})\rho_{12}Y_A
\frac{2}{7} N_p(Z)N_h(N) T_{11}^6  
F_5^{q'-}(\eta_{\nu_e},\eta_{e})$&   \\
5 &$n+n \to n+p+e+\nu_e$ &$q_{URCA}(\nu_e)=2.7\times 
10^{34}({\rm erg cm}^{-3}{\rm sec}^{-1})\rho_{12}^{2/3}T_{11}^8
F_Q(\eta_e,\eta_{\nu_e}) $ & Shapiro et al.(1983) \\ 
6 &$\gamma \to \nu_i + \bar\nu_i$ & $q_\gamma(\nu_e,\bar{\nu}_e) 
\approx 1.20\times 10^{33}({\rm erg cm}^{-3}{\rm sec}^{-1}) T^9
F_{Q\gamma}(\eta_{\nu_e})$ &Ruffert et al.(1996)  \\ 
  &  &$q_\gamma(\nu_x) \approx 8.3\times 10^{30}({\rm erg cm}^{-3}{\rm sec}^{-1}) T^9 
F_{Q\gamma}(\eta_{\nu_x})$ &Ruffert et al.(1996)\\
7 &$e+e^+ \to \nu_i+ \bar{\nu}_i $&$q_{ee}(\nu_e,\bar{\nu}_e) \approx 1.24\times 
10^{33}({\rm erg cm}^{-3}{\rm sec}^{-1})T_{11}^9f_{\nu_e}(\eta)$  &  Thompson et al.(2000) \\
 &  &$q_{ee}(\nu_x,\bar{\nu}_x) \approx 5.49\times 10^{32}({\rm erg cm}^{-3}{\rm sec}^{-1})T_{11}^9f_{\nu_x}(\eta) 
$ & Thompson et al.(2000)  \\
\hline
\end{tabular}
\footnote{The emission rates $q$ are expressed with unit
 erg/cm$^3$/sec. The complex integrals, $F_{5}^-(\eta_{\nu_e},\eta_e,q), 
F_{5^+}^-(\eta_{\bar{\nu}_e},\eta_{e^+},q)$ and $F_5^{q'}(\eta_{\nu_e})$, are 
shown in Table A5. $F_Q$ and $f_{\nu_i}$ for $i=e,x$ are 
$F_Q(\eta_e,\eta_{\nu_e})=<1-f_e(\varepsilon_e)><1-f_{\nu_e}(\varepsilon_{\nu_e}
)>$ and 
$f_{\nu_i}(\eta)=\frac{F_4(\eta_e)F_3(\eta_{e^+})+F_4(\eta_{e^+})F_3(\eta_e)}{2F
_4(0)F_3(0)} <1-f_{\nu_i}(\varepsilon)>_{ee}<1-f_{\bar\nu_i}>_{ee}$. 
$F_{Q\gamma}(\eta_{\nu_e})$ is $F_{Q\gamma}(\eta_{\nu_e})=\gamma^6 
e^{-\gamma}\frac{1}{2} ((1+(\gamma+1)^2)<1-f_{\nu_e}(\varepsilon)>_\gamma 
<1-f_{\bar{\nu}_e}(\varepsilon)>_\gamma$, where $\gamma$ and $\gamma_0$ are  
$\gamma\approx \gamma_0\sqrt{\pi^2/3+\eta_e^2}, \quad \gamma_0=h 
\Omega_0/(2\pi m_e c^2) =5.565\times 10^{-2}$. The mean blocking factors 
$<1-f_{\nu_i}(\varepsilon)>_{ee}, ..$ are described in Table A5. The
density $\rho_{12}$ and temperature $T_{11}$ are normalized vales 
by $10^{12}$(g/cm$^3$) and $10^{11}$K}. 
\end{minipage}
\end{table*}

\begin{table*}
\begin{minipage}{160mm}
\caption{Fraction Change}
\label{tbl-5}
\begin{tabular}{@{}lll}
No. & {Neutrino Reaction} & {Fraction Change} \\
\hline
1 &$\nu_e + n  \rightleftharpoons e + p$ &$\dot{Y}_e(\beta)=
\frac{4\pi c} {(h c)^3} \int_0^\infty d^3p\sigma_{\nu_e, 
n}^a(\varepsilon)\Bigr[ Y_{np}  f_{\nu_e}(\varepsilon) [1-f_{e}(\varepsilon+Q)] - 
Y_{pn} f_{e}(\varepsilon+Q) [1-f_{\nu_e}(\varepsilon)]\Bigr]$ \\ 
  &  &$\approx  9.0\times 10^2 ({\rm sec}^{-1})
T_{11}[Y_{np}F_4^+(\eta_{\nu_e},\eta_e,q)-Y_{pn}F_4^-(\eta_{\nu_e},\eta_e,q)]$
 \\ 
2 &$\nu_e + A \rightleftharpoons e+A'$ &$\dot{Y}_e(A)=b\frac{\alpha^2}{4}\sigma_0 
Y_A \frac{2}{7} N_p(Z) N_h(N)\Bigl(\frac{kT}{m_e c^2} \Bigr)^5 \big(e^{\eta_n 
-\eta_p -q'} F_4^{q'+}(\eta_{\nu_e}, \eta_e) - F_4^{q'-}(\eta_{\nu_e}, \eta_e) 
\big)$ \\
   &$    $&$\approx 2.4\times 10^{2} ({\rm sec}^{-1}) T_{11}^5  Y_A \frac{2}{7} 
N_p(Z)N_h(N)\Bigl(e^{\eta_n -\eta_p -q'} F_4^{q'+}(\eta_{\nu_e},\eta_e)  - 
F_4^{q'-}(\eta_{\nu_e}, \eta_e) \Bigr)$ \\
3 &$n+n \to n+p+e+\nu_e $ &$\dot{Y}_e(URCA)\approx 
\frac{q_{URCA}}{\bar{\varepsilon}_{\nu_e} n_B}$ \\
  & &$\approx
3.3\times 10^{3} ({\rm sec}^{-1})\rho_{12}^{-1/3}T_{11}^7 
\frac{F_2(\eta_{\nu_e})}{F_3(\eta_{\nu_e})} <1-f_e(\varepsilon)>_{nn} 
<1-f_{\nu_e}(\varepsilon)>_{nn}$ \\
4 &$\bar{\nu}_e + p \rightleftharpoons  e^+ + n$&$\dot{Y}_{e^+}(\bar{\nu}_e p)
\approx 9.0\times 10^2 ({\rm sec}^{-1})
T_{11}[Y_{pn}F_4^+(\eta_{\bar{\nu}_e},\eta_{e^+},-q)-Y_{np}F_4^-(\eta_{{\bar\nu}
_e},\eta_{e^+},-q)]$\\
5 &$N+N \to N+N+\nu_i+\bar{\nu}_i $&$\dot{Y}_{\nu_i}(NN)\approx 
\frac{q_{\nu\bar\nu}(nn)}{2\bar{\varepsilon}_{\nu_i}n_B} \approx 
1.3 ({\rm sec}^{-1})
X_N^2\rho_{12}T_{11}^5\frac{F_2(\eta_{\nu_i})}{F_3(\eta_{\nu_i})}\Bigl(\frac{3}{
2+0.862T_{11}}\Bigr)^{1/2} $ \\
6 &$e+e^+ \rightleftharpoons \nu_i+\bar{\nu}_i$ &$\dot{Y}_{\nu_e}(ee^+)\approx 
36.3 ({\rm sec}^{-1})T_{11}^8 f_{3,\nu_e}(\eta)$ \\
 &  &$\dot{Y}_{\nu_x}(ee^+) \approx 31.2 ({\rm sec}^{-1})T_{11}^8 f_{3,\nu_x}(\eta)   $
 \\
7 &$\gamma \to \nu_i + \bar\nu_i $&$\dot{Y}_{\nu_e}(\gamma)\approx 1.4\times 10^2({\rm sec}^{-1}) 
F_\gamma(\eta)\rho_{11}^{-1} T^8_{11}  $\\
  &  &$\dot{Y}_{\nu_x}(\gamma)\approx 1.0 ({\rm sec}^{-1}) F_\gamma(\eta)\rho_{11}^{-1} T^8_{11} 
$\\
\hline
\end{tabular}
\footnote{The parameter $b$ is $b=4\pi \Bigl(\frac{m_ec}{h}\Big)^3$. 
The complex integrals, $F_4^{q'+}(\eta_{\nu_e}, \eta_e)$
 and $F_4^{q'}(\eta_{\nu_e}, \eta_e)$, are shown in Table
 A5.   $f_{3,\nu_i}(\eta)$ and $F_{Q\gamma}(\eta_{\nu_e})$ are 
$f_{3,\nu_i}(\eta)=(F_3(0))^{-2}\Bigl\{ F_3(\eta_e)F_3(\eta_{e^+}) 
<1-f_{\nu_i}(\varepsilon)>_{ee}<1-f_{{\bar\nu}_i}(\varepsilon) >_{ee} 
- F_3(\eta_{\nu_i})F_3(\eta_{{\bar\nu}_i}) <1-f_{e}(\varepsilon)>_{\nu_i 
{\bar\nu}_i}<1-f_{e^+}(\varepsilon)>_{\nu_i {\bar\nu}_i}  \Bigr\}$, and 
$F_{Q\gamma}(\eta_{\nu_e})=\gamma^6 e^{-\gamma}\frac{1}{2} 
((1+(\gamma+1)^2)<1-f_{\nu_e}(\varepsilon)>_\gamma 
<1-f_{\bar{\nu}_e}(\varepsilon)>_\gamma$. The mean blocking factors 
$<1-f_{\bar{\nu}_e}(\varepsilon)>_\gamma, ...$ are described in Table 5.  $X_N$ is the mass fraction 
of neucleons.  }
\end{minipage}
\end{table*}

\begin{table*}
\begin{minipage}{160mm}
\caption{Integrals and Blocking Factors}
\label{tbl-6}
\begin{tabular}{@{}ll}
No. & {Integrals and Blocking Factors} \\
\hline
1 & $F_{k}^+(\eta_{\nu},\eta_{e},q) = \int_0^\infty 
\frac{x^{k-2}\Theta(x+q-m)(x+q)^2} 
{1+\exp{(x-\eta_{\nu})}}(1-f_e(q+x))dx=\int_0^\infty 
\frac{x^{k-2}\Theta(x+q-m)(x+q)^2dx} 
{(1+\exp{(x-\eta_{\nu})})(1+\exp{(\eta_{e}-q-x)})}$ \\
2 & $F_{k}^-(\eta_\nu,\eta_e,q)=\int_0^\infty \frac{x^{k-2}\Theta(x+q-m)(x+q)^2}
{(1+\exp{(x+q-\eta_e)})}(1-f_\nu(x))dx = \int_0^\infty 
\frac{x^{k-2}\Theta(x+q-m)(x+q)^2dx} {(1+\exp{(x+q-\eta_e)}) 
(1+\exp{(\eta_\nu-x)})}$\\
3 & $F_k^{q'-}(\eta_{\nu}, \eta_e) = \int_0^\infty \frac{(x+q')^2 
x^{k-2}}{\exp{(x+q'-\eta_e)} +1}(1-f_{\nu}(x))dx = \int_0^\infty \frac{(x+q')^2 
x^{k-2} dx}{(\exp{(x+q'-\eta_e)} +1)(\exp{(\eta_{\nu}-x)} +1)}$  \\
4 & $F_k^{q'+}(\eta_{\nu},\eta_e) = \int_0^\infty 
\frac{x^{k-2}(x+q')^2}{1+\exp{(x-\eta_{\nu})}}(1-f_e(q'+x))dx = \int_0^\infty 
\frac{x^{k-2}(x+q')^2 dx} {(1+\exp{(x-\eta_{\nu})}) (1+\exp{(\eta_e -q' 
-x)})}$  \\
5 & $<1-f_{e}(\varepsilon)>_{nn} \approx 
\Bigl\{1+\exp{[\eta_{\nu}-\frac{F_5(\eta_n}{F_4(\eta_n)}]} \Bigr\}^{-1}$ \\
6 & $<1-f_{\nu}(\varepsilon)>_{nn} \approx 
\Bigl\{1+\exp{[\eta_{\nu}-\frac{F_5(\eta_n}{F_4(\eta_n)}]} \Bigr\}^{-1}$ \\
7 & $<1-f_{\nu_i}(\varepsilon)>_{ee} \approx 
\Bigl\{1+\exp{\Bigl[\eta_{\nu_i}-\frac{1}{2}\Bigl(\frac{F_5(\eta_e)}{F_4(\eta_e)
}+\frac{F_5(\eta_{e^+})}{F_4(\eta_{e^+})}\Bigr) \Bigr]} \Bigr\}^{-1}$ \\
8 & $<1-f_{e}(\varepsilon+q')> \approx \Bigl\{1+\exp{\big[ \eta_e 
-\frac{F_5(\eta_{\nu_e})}{F_4(\eta_{\nu_e})} -q'\big]} \Bigr\}^{-1}$  \\
9 & $<1-f_{\nu_i}(\varepsilon)>_\gamma \approx 
\Bigl(1+\exp{\big[\eta_{\nu_i}-(1+\frac{1}{2}\frac{\gamma^2}{1+\gamma})\big]}\Bigr)^{-1}$  \\
\hline
\end{tabular}
\footnote{ The quantity $\Theta(x)$ is the unit step function,
 $\Theta(x)=0$ if $x<0$, and  $\Theta(x)=1$ if $x \ge 0$.  }
\end{minipage}
\end{table*}


\section[]{Transfers of neutrinos in a plane-parallel disk}
The analytic formulas of neutrino's transfers are derived by using the
simplified treatment of radiative transfer given by
\citet{PN95}. Introducing the two streams, with intensities $I^+, I^-$
moving at an angle  $\cos\theta =1/\sqrt{3}$
relative to $+z$ and $-z$, define the net flux density $j$ and the
total intensity $\bar{j}$ such as ${j}=(I^+ - I^-)/2\sqrt{3}$ and $ \bar
j=(I^+ + I^-)/2$ \citep{PN95}. 
The transfers of neutrinos are then described by the production rate of
neutrinos, $\dot{n}^+$, and the absorption opacity, scattering opacity
and the total opacity,
$\kappa_{a}, \kappa_{s}$ and $\kappa=\kappa_{a}+\kappa_{s}$, as  
\begin{eqnarray}
\frac{d{j}}{dz}&=&- \kappa_a\bar j + \dot{n}^+ \label{I1}\\
\frac{d\bar j}{dz} &=&- 3\kappa {j} \label{I2}.
\end{eqnarray}
 
The boundary condition at the center of the disk is
$j=0$. To derive the simplest solution, let us assume that $(\dot{n}^+
-\kappa_a\bar j), \kappa$ are independent of $z$ \citep{Hub}. Equations (\ref{I1}) and (\ref{I2}) can then be integrated to
give 
\begin{eqnarray}
j(z)&=&(\dot{n}^+_c - \kappa_a\bar{j}_c)z \label{J3}\\
\bar j(z) &=&\bar{j}_c - \frac{3}{2}\kappa(\dot{n}_c^+ - \kappa_a\bar{j}_c)z^2 
\label{J4}, 
\end{eqnarray}
where $\dot{n}^+_c$ and $\bar{j}_c$ are the values of $\dot{n}^+$ and
$\bar{j}$ at $z=0$. The boundary condition at the surface of the disk is $j^-=0$, 
i.e., $\bar
j=\sqrt{3}{j}$. Applying this boundary condition we find  
\begin{equation}
\bar j_c=\frac{\sqrt{3}+1.5\tau}{1+\sqrt{3}\tau_a+1.5\tau\tau_a}\dot{n}_c^+H,
\end{equation}
where $\tau_a$ is the absorbing optical depth, $\tau_a=\kappa_aH$, and
$\tau$ is the total optical depth, $\tau=\kappa H$. 
The escaping flux density of any sort of neutrino, $\nu_i$, is then given by  
\begin{equation}
F_{n(\nu_i)}={j}(H)=\frac{\dot{n}^+_{\nu_i}H}{1.5\tau_{\nu_i 
a}\tau_{\nu_i}+\sqrt{3}\tau_{\nu_i a} + 1},   
\end{equation}
where $\tau_{\nu_i}$ and $\tau_{\nu_i a}$ are the total optical depth
and the absorbing one of $\nu_i$, and $\dot{n}^+_{\nu_i}$ is the production rate
of $\nu_i$ per unit volume.     

\section{Changing Rates of Positrons along the Flow}
The number density of positrons which are equilibrium with thermal photons 
is described as
\begin{equation}
n_{e^+th} =
\frac{1.8}{\pi^2}\Bigl(\frac{kT}{hc/2\pi}\Bigr)^3<1-f(\eta_e)><1-f(\eta_{e^+})>. 
\end{equation}
(Landau and Lifshitz 1964), where the brackets indicate the mean values of 
the phase space blocking factors. These factors are evaluated with the average energy of electrons and positrons produced by $e^-e^+$-pair creation\citep{RJS}:
\begin{equation}
<1-f(\eta_{e^+})> \cong \Bigl\{1+\exp{[-(\bar{\varepsilon}^+ - \eta_{e^+})]} 
\Bigr\}^{-1} .
\end{equation}
The average energies, $\bar{\varepsilon}^+$ and $\bar{\varepsilon}^-$, are approximately 
given by 
\begin{eqnarray}
\bar{\varepsilon}^+=\bar{\varepsilon}^- 
=\frac{\varepsilon_{e^+}}{n_{e^+}}=\frac{7\pi^4}{180\zeta(3)}kT,
\end{eqnarray}
where $\varepsilon_{e^+}$ and $n_{e^+}$ are the energy and number 
densities of positrons. When the fraction of positrons becomes large, $Y_{e+} 
\ge 0.01$, the changing rate of positron number, ${\dot n}_{e^+}$, due
to the reactions, $e^+ + n \rightleftharpoons \bar{\nu}_e + p \quad$,
$e+e^+\rightleftharpoons 
\nu_i+\bar{\nu}_i$, cannot be negligible. We calculate the changing rate of the positrons along the flow by using the following equation:
\begin{eqnarray}
\dot{Y}_{e^+}=n_b^{-1}v^r\nabla_r n_{e^+}= n_b^{-1}(\dot{n}_{e^+} + v^r\nabla_r {n_{e^+ th}}),
\end{eqnarray}
where $\dot{n}_{e^+}$ is the net production rate of positrons,
$\dot{n}_{e^+} =  -\dot{n}_{\bar\beta} -\dot{n}_{e e}$.

\section{Dynamical Structure of the Disk}
Dynamically equilibrium structure of the disk is given by the
Euler equation. The equation of motion of the fluid, $h^k_iT^{j}_{k;j}=0$,
for the steady, axially symmetric and rotating ideal fluid can be
expressed in the total differential form:
\begin{equation}
\frac{dp}{\rho+\epsilon+P}=\gamma^2(-d\nu+v_{(\varphi)}^2d\psi-v_{(\varphi)}v_\omega d\ln\omega) \equiv dU. \label{eq dpA}
\end{equation}
The equipressure surfaces are given by the equation $U=$constant. The
quantity $U=U(p)$ is equal in the Newtonian limit to the total potential
(gravitational plus centrifugal ones) expressed in the units of $c^2$
\citep{AJS78}. 
 We solve the disk structure in terms of cylindrical coordinates $
 \varpi, z,$ and $\varphi$ instead of radial coordinates which are
 combined through the relations,
\begin{eqnarray}
r \equiv \sqrt{\varpi^2+z^2}, \quad  \theta \equiv \cos^{-1}{(z/r)}.
\end{eqnarray}
If no pressure force exerts in the $\varpi$ direction, the equation
$(\ref{eq dpA})$ determines the velocity of the geodesic circular orbit,
$v_{(\varphi)}$. Let's introduce the new variables,
$\tilde{a},\tilde{m},s, c$ defined by $\tilde{a}\equiv a/r,
\quad\tilde{m}\equiv M/r$, $s \equiv 
\sin{\theta}, \quad c \equiv \cos{\theta}$. The following functions
 are defined by: 
\begin{eqnarray}
&\Sigma' &= 2rs(1- \tilde{a}^2c^2), \quad \Delta' = 2rs(1-\tilde{m}) \nonumber  \\
&A' &= 2r^3s \Bigl\{ 2+\tilde{a}^2(1+\tilde{m}(1+c^2)- \tilde{a}^2c^2) \Bigr\}, \nonumber \\
&A_v &= 2\psi_{, \varpi} = \frac{A'}{A} - \frac{\Sigma'}{\Sigma} + \frac{2c^2}{rs}, \nonumber \\
&B_v &= -v_\omega \frac{\partial\ln{\omega}}{\partial \varpi} = \frac{s}{r} - \frac{A'}{A}, \nonumber \\
&C_v &= 2\nu_{, \varpi}= \frac{\Sigma'}{\Sigma} +\frac{\Delta'}{\Delta} -\frac{A'}{A}.\nonumber
\end{eqnarray}
The circular velocity is then given by
\begin{equation}
v_{(\varphi)} = \frac{-B_v \pm \sqrt{B_v^2+A_vC_v}}{A_v}. \label{v-varphi}
\end{equation}
This velocity at the equatorial plane is expresses as
\begin{eqnarray}
v_{(\varphi)} = \frac{\Theta_1(r)}{r^{1/2}},\quad
\Theta_1(r) \equiv \frac{1-2aM^{1/2}r^{-3/2} + a^2r^{-2}} 
{(1- 2Mr^{-1}+ a^2r^{-2})^{1/2} (1+aMr^{3/2})}.  \label{Psi1}
\end{eqnarray}

The vertical force is described by the gradient of the equipotential,
$\nabla_z U$. The first order of the expansion of $\partial U/\partial z$ in
$z/r ( \ll 1)$ gives the approximate expression such as  
\begin{eqnarray}
&\frac{\partial U}{\partial z} &\approx -\frac{Mz}{r^3}\Phi, \nonumber \\
&\Phi(r) &=
\frac{1-\omega/\Omega}{D}\big(\frac{\gamma\Psi}{\Theta}\big)^2
\biggl\{ 1 + \frac{1+v_{(\varphi)}^2}{v_{(\varphi)}v_\Omega} \big[
1+(\frac{a}{r})^2 \big]  \nonumber \\
&-& (1+ \frac{  \omega}{\Omega} + \frac{1}{v_{(\varphi)} v_\Omega})\frac{\acute{A}}{\Psi} + \frac{1-M/r}{v_{(\varphi)} v_\Omega D}  \biggr\},  \nonumber \\
&\Theta(r)&=1+ \frac{aM^{1/2}}{r^{3/2}}, \quad D=1-\frac{2M}{r} +
\big(\frac{a}{r}\big)^2,\nonumber \\
&\Psi &= 1 + \big(\frac{a}{r}\big)^2\big(1 + 2\frac{M}{r}\big),  \nonumber \\ 
&\acute{A} &= 2 + (4- 3\frac{M}{r})\big(\frac{a}{r}\big)^2 +
\big(\frac{a}{r}\big)^4, \nonumber \\ 
 &v_\Omega &= \Omega e^{\psi-\nu}=v_{(\varphi)} + \omega e^{\psi-\nu}   \label{Psi}
\end{eqnarray}


\label{lastpage}

\end{document}